\documentclass[lettersize,journal]{IEEEtran}
\usepackage{amsmath,amsfonts}
\usepackage{amssymb}
\usepackage{algorithmicx}
\usepackage{algorithm}
\usepackage{algpseudocode}
\usepackage{array}
\usepackage{textcomp}
\usepackage{stfloats}
\usepackage{url}
\usepackage{verbatim}
\usepackage{graphicx}
\usepackage{cite}
\usepackage{subfig}
\usepackage{colortbl,booktabs}
\newtheorem{assumption}{Assumption}
\newtheorem{theorem}{Theorem}
\newtheorem{corollary}{Corollary}
\newtheorem{lemma}{Lemma}

\newtheorem{remark}{Remark}
\begin{document}

\title{Supplementary material – Approximate Dynamic Programming for Constrained Piecewise Affine Systems with Stability and Safety Guarantees}

\author{Kanghui He$^1$, Shengling Shi$^2$, Ton van den Boom$^1$, and Bart De Schutter$^1$% <-this % stops a space
	\thanks{This paper is part of a project that has received funding from the European Research Council (ERC) under the European Union’s Horizon 2020 research and innovation programme (Grant agreement No. 101018826 - CLariNet).}% <-this % stops a space
	\thanks{$^1$Kanghui He, Ton van den Boom, and Bart De Schutter are affiliated with Delft Center for Systems and Control, Delft University of Technology, Delft, The Netherlands (e-mail: \{k.he, a.j.j.vandenboom, b.deschutter\}@tudelft.nl). $^2$Shengling Shi is affiliated with Department of Chemical Engineering, Massachusetts Institute of Technology, Cambridge, Massachusetts, USA (e-mail: slshi@mit.edu).}%
}

\markboth{Journal of \LaTeX\ Class Files,~Vol.~14, No.~8, April~2023}%
{Shell \MakeLowercase{\textit{et al.}}: A Sample Article Using IEEEtran.cls for IEEE Journals}


\maketitle

\begin{abstract}
	
Infinite-horizon optimal control of constrained piecewise affine (PWA) systems has been approximately addressed by hybrid model predictive control (MPC), which, however, has computational limitations, both in offline design and online implementation. In this paper, we consider an alternative approach based on approximate dynamic programming (ADP), an important class of methods in reinforcement learning. We accommodate non-convex union-of-polyhedra state constraints and linear input constraints into ADP by designing PWA penalty functions. PWA function approximation is used, which allows for a mixed-integer encoding to implement ADP. The main advantage of the proposed ADP method is its online computational efficiency. Particularly, we propose two control policies, which lead to solving a smaller-scale mixed-integer linear program than conventional hybrid MPC, or a single convex quadratic program, depending on whether the policy is implicitly determined online or explicitly computed offline. We characterize the stability and safety properties of the closed-loop systems, as well as the sub-optimality of the proposed policies, by quantifying the approximation errors of value functions and policies. We also develop an offline mixed-integer linear programming-based method to certify the reliability of the proposed method. Simulation results on an inverted pendulum with elastic walls and on an adaptive cruise control problem validate the control performance in terms of constraint satisfaction and CPU time.
	
\end{abstract}
\begin{IEEEkeywords}
	Reinforcement learning, approximate dynamic programming, constrained control, piecewise affine systems.
\end{IEEEkeywords}

%%%%%%%%%%%%%%%%%%%%%%%%%%%%%%%%%%%%%%%%%%%%%%%%%%%%%%%%%%%%%%%%%%%%%%%%%%%%%%%%
\section{INTRODUCTION}
\subsection{Backgrounds}
There has been an increasing interest in control of piecewise affine (PWA) systems due to their capability of representing hybrid models and approximating nonlinear dynamics \cite{borrelli2017predictive}. Many practical control problems can be modelled as PWA systems with constraints, including emergency evasive maneuvers \cite{gharavi2023efficient}, robotic manipulation that has multi-contact behaviors \cite{sadraddini2019sampling}, and traffic control \cite{groot2012integrated}. PWA systems are a special class of switched systems where the subsystem in each mode is affine and the transitions are based on the state belonging to different regions. Tractable controller design methods for PWA systems include synthesizing piecewise linear control laws via Linear Matrix Inequalities \cite{lazar2006stabilizing}, adaptive control \cite{liu2021adaptive}, and model predictive control (MPC) \cite{lazar2006stabilizing}. \textcolor{blue}{The challenges of controlling a PWA system include ensuring stability and achieving optimality guarantees \cite{lazar2006stabilizing}, as well as addressing both offline and online computational complexity \cite{borrelli2005dynamic,sadraddini2019sampling}. These difficulties primarily arise from the system's hybrid structure and inherent nonlinearity.} For suboptimal control of PWA systems with constraints, MPC is widely applied. However, MPC for PWA systems still faces challenges in computational complexity \cite{borrelli2005dynamic}, because it involves solving a mixed-integer linear programming (MILP) problem. The complexity of solving MILP MPC problems is in general dominated by the number of integers, which is proportional to the prediction horizon. Explicit MPC \cite{borrelli2005dynamic}, an offline version of MPC, requires solving a parametric MILP problem, which is also suffering from computational complexity issues. These issues make MPC only suitable for slow PWA processes or for low-dimensional problems with short prediction horizons \cite{sadraddini2019sampling}. %Methods aimed at alleviating computational burden include decision trees \cite{masti2020learning,sadraddini2019sampling}, optimistic optimization \cite{xu2016model}, warm start of MPC \cite{marcucci2020warm}, etc.

In contrast to MPC, reinforcement learning (RL) can learn a policy that minimizes a finite-/infinite-horizon cost and has a much lower online computational burden \cite{gorges2017relations} than MPC. In RL, two different methodologies can be distinguished: policy search and dynamic programming \cite{busoniu2017reinforcement}. Policy search directly searches for an optimal policy, while dynamic programming aims to solve the Bellman equation first and then determines the policy. Dynamic programming has the advantage over policy search is that it reduces the policy optimization problem to a one-step look-ahead problem. When applied to systems with continuous state and input spaces, %dynamic programming needs to solve an infinite number of optimization problems. To deal with this issue, 
approximate dynamic programming (ADP), which uses sample-based methods accompanied by sophisticated function approximators, has been developed \cite{heydari2014revisiting,wei2015value}. In this paper, we consider approximate value iteration (VI), the most basic and direct way to solve the Bellman equation. Moreover, we provide comprehensive performance guarantees for stability, safety and sub-optimality of the developed ADP approach. In the context of RL, safety can have various definitions. In this paper, we specifically address safety as ensuring that the state and input of the system satisfy predefined constraints throughout the entire system's evolution after the learning process is finished.

\subsection{Related work}
To reduce the computational cost of MPC, two different types of approaches have been extensively studied: approximate MPC and reinforcement learning under constraints.

\emph{Approximate MPC:} Approximate MPC parameterizes a policy and then uses supervised learning or gradient-based methods to mimic a predictive control policy. The online computational cost is thus significantly reduced because approximate MPC directly computes control actions based on the learned parameters, rather than online solving an optimization problem. Some approximate MPC work considers linear systems, with different focuses on, e.g., stability \cite{schwan2022stability} and constraint satisfaction \cite{chen2018approximating}. Some approximate MPC approaches can handle nonlinear control problems with constraints, e.g., by using constraint tightening \cite{hertneck2018learning}. However, as these approaches are developed for general nonlinear systems, they lack deterministic performance verification tools.
	
\emph{RL for constrained control:} RL for constrained control can be roughly categorized into two groups: policy-projection RL methods, which impose constraints \emph{after} learning, and policy-optimization RL methods, which impose constraints \emph{during} learning. Using RL for controller synthesis allows designers to stop the learning process and get a satisfactory policy whenever the expected performance is met. %Compared to directly approximating an MPC controller, which is usually derived from a fixed problem, RL can be more adaptive in the sense of its inherent iterative framework. 
	
	For the first group, a predictive safety filter \cite{wabersich2021predictive,li2022robust} can be adopted to modify the learned policy, which can be derived from any RL algorithm. This group of approaches can in general be applied to constrained nonlinear control problems. For PWA systems with linear constraints, this group needs to solve mixed-integer convex problems online \cite{li2022robust}, and as a result it is not suitable for large-scale systems or for systems requiring fast computation. %In \cite{li2022robust}, the probably unsafe policy is adjusted by a proposed action governor, which needs to compute and store the maximal controlled-invariant set. However, computing this set is intractable for high dimensional systems \cite{anevlavis2019computing}. 
	Additionally, this group of methods in general jeopardizes optimality and stability \cite{gros2020safe}.
	
	For the second group, constrained policy optimization \cite{chakrabarty2019approximate,beckenbach2018constrained,duan2022adaptive} is often used. Most methods \cite{tessler2018reward} consider constrained Markov Decision Process (MDP) problems, in which constraints are on expected cumulative costs. Recent developments have been made to transform instantaneous constraints into constraints on expected cumulative costs. Among these developments, \cite{li2021augmented} considers instantaneous expectation constraints, while \cite{paternain2019learning} contemplates chance constraints. In \cite{yu2022reachability}, hard instantaneous constraints are considered, but the convergence of the RL algorithm relies on the assumptions of the finiteness of MDP and the continuous differentiability of the value function, which is not the case for optimal control of PWA systems. Besides, \cite{yu2022reachability,li2021augmented} do not provide stability or safety analysis of the RL policy.

	\emph{Lyapunov stable RL:} In addition to the online computational advantage, the stability property of RL controllers has been investigated recently. In \cite{heydari2018stability}, stability analysis of approximate VI for unconstrained nonlinear systems is conducted, but it results in very restrictive conditions, which require the approximation errors per iteration to be upper bounded. \cite{postoyan2019stability,moreno2022predictive} further relax these conditions by only focusing on the stopping criterion and the approximation error in the last iteration. %It is shown that, under mild assumptions on the stabilizability, detectability, as well as the boundedness of the approximation errors in the final iteration, the closed-loop system exhibits stability properties. 
	Nevertheless, these references consider unconstrained problems and do not address the sub-optimality of the RL policy. Besides, the combination of RL and MPC is proposed in \cite{gros2022learning,moreno2022predictive,yang2021data,he2022approximate} as a theoretically justified methodology to produce safe and stabilizing policies. However, in \cite{gros2022learning,moreno2022predictive,yang2021data,he2022approximate}, the online computational issue of MPC is not solved by introducing RL. 
	
	Based on these observations, for the policy optimization methods, one could conclude that (i) no work has been done yet for PWA systems, and (ii) good and comprehensive performance of RL-based controllers regarding online computing convenience, stability, and constraint fulfillment cannot be achieved simultaneously. In this paper, we will consider policy optimization methods for PWA systems subject to linear or union-of-polyhedra (UoP) constraints, and solve the above two issues.
	
	 Meanwhile, most RL/ADP methods ignore the properties of the value function and the policy, e.g., boundedness, continuity, nonlinearity, so that the design of function approximators is usually done in an empirical way \cite{bucsoniu2018reinforcement}. For PWA systems with a linear performance index, fortunately, extant research on explicit hybrid MPC has shown that the optimal value function and the optimal policy are both PWA \cite{baoti2006constrained,borrelli2017predictive}. This provides an opportunity to use PWA function approximation, as will be investigated in this paper.

	%Besides, in \cite{berkenkamp2017safe}, a stable model-based RL is proposed by adding a reward term for the decrease of the candidate Lyapunov function during the dynamic programming process. However, \cite{berkenkamp2017safe} lacks theoretical results on the convergences.
\emph{Performance verification of learning-based controllers:} In addition to controller design, there is some related work on performance analysis and verification of RL or any learning-based controllers. In \cite{chen2021learning,dai2021lyapunov}, a learner/verifier framework is proposed to verify offline the asymptotic stability for an input-constrained PWA system controlled by a PWA neural network (NN). Using MILP to explicitly compute the set of trajectories of linear systems controlled by an NN can also be applied to evaluate the safety and stability \cite{karg2020stability}. However, for nonlinear systems with NN controllers, achieving practical stability is much easier than achieving asymptotic stability \cite{postoyan2019stability}. For PWA systems with NN controllers, there is currently no systematic way to verify different properties including practical and asymptotic stability as well as state constraint satisfaction. 

%\textcolor{blue}{In \cite{baoti2006constrained}, an exact value iteration algorithm is designed for optimal control of PWA systems, by solving a series of multi-parametric linear programs (mp-LPs). It is shown in \cite{baoti2006constrained} that when the algorithm starts from a zero value function, closed-loop stability can be achieved only if the algorithm terminates at the optimal value function. %This makes the algorithm rarely tractable in practice because finite-time termination is not guaranteed. Besides, the number of mp-LPs needed to solve as well as the number of polyhedral regions that the optimal policy has can grow exponentially with the system dimension.}
\subsection{Methods and contributions of this paper}

In conclusion, using RL to produce a reliable learning-based controller for constrained PWA systems with performance guarantees and low online computational requirements is still an open problem. The main challenge is to concurrently ensure the stability, safety, and efficiency of the online computations. Existing work can either provide stability/safety guarantees \cite{wabersich2021predictive,li2022robust,gros2022learning,moreno2022predictive,he2022approximate,beckenbach2018constrained} or achieve low computational cost \cite{yu2022reachability,li2021augmented}. In this paper, we develop ADP algorithms under linear and UoP constraints. We propose two formulations for the inclusion of PWA penalties in dynamic programming, i.e., adding penalties to the stage cost and integrating penalties into the cost-to-go. For exact dynamic programming algorithms, we prove the continuity and the PWA property of the value functions when PWA penalty functions and UoP constraints are considered. These properties allow us to design PWA function approximators. We then present two different controllers: an implicit controller that is obtained online by solving an MILP problem that is much more simple than the one of implicit hybrid MPC, and an explicit controller that is learned offline by policy gradient. The explicit policy is then projected online onto the input constraint set, which corresponds to solving a single convex quadratic program online. 

We provide rigorous analysis on the closed-loop stability, safety, as well as sub-optimality of the controllers. We establish a systematic, MILP-based procedure that allows us to certify the reliability, in terms of stability and safety of the closed-loop system. 

The paper contributes the state of the art as follows
 
 \begin{enumerate}
 	\item This work is the first research on designing policy optimization RL methods for constrained PWA systems. Systematic performance analysis on the feasibility, stability, and sub-optimality of the RL-based controllers is provided. The analysis suggests several ways to employ the proposed algorithms in practice.
 	
 	\item Compared to MPC, our method exhibits a superiority in terms of online computational simplicity. In particular, the resulting online policy optimization problem is either an MILP problem with significantly fewer integer variables than the hybrid MPC MILP problem, or a single convex quadratic programming (QP) problem.%Simulation results show that the proposed controllers perform better in state constraint satisfaction and the level of sub-optimality, compared to the method in \cite{wabersich2021predictive}.

 	\item We develop a mixed-integer optimization-based framework to exactly verify the stability and safety of the closed-loop system. The framework extends the verification techniques of \cite{chen2021learning,dai2021lyapunov,karg2020stability,schwan2022stability}. Specifically, we establish a comprehensive scheme that addresses both practical and asymptotic stability properties and the enlargement of stable and safe regions while \cite{chen2021learning,dai2021lyapunov,karg2020stability,schwan2022stability} only investigate asymptotic stability. %We consider PWA dynamics with UoP state constraints while \cite{karg2020stability,schwan2022stability} focus on linear systems and \cite{chen2021learning,dai2021lyapunov} consider PWA systems with polytopic constraints.
 	
 	\item For constraints handling, we explore how the proposed two manners of incorporating PWA penalties can accelerate the learning process. We discuss their adaptability in different situations.

 \end{enumerate}

%%%%%%%%%%%%%%%%%%%%%%%%%%%%%%%%%%%%%%%%%%%%%%%%%%%%%%%%%%%%%%%%%%%%%%%%%%%%%%%%
\section{Preliminaries}

\emph{Notations:}  Let $\mathbb{R} = (-\infty,+\infty)$, $\mathbb{R}_{\geq 0} = [0, +\infty)$, and $\mathbb{R}_{> 0} = (0, +\infty)$. The boundary of the set $ \mathcal{S}$ is $\partial \mathcal{S}$, and $\mathrm{int}(\mathcal{S})$ stands for the interior of $ \mathcal{S}$. We utilize $A_{i\cdot}$ to represent the $i$th row of the matrix $A$. %The operator $\oplus$ represents the Minkowski set addition, and $A \mathcal{S}$ returns a set $\{A x| x \in \mathcal{S}\}$ for the set $\mathcal{S}$ and the matrix $A$. 
We define the sub-level set $\mathcal{B}(J,S)$ for a continuous function $J: \mathbb{R}^n \to \mathbb{R}_{\geq 0}$ and a compact set $S \subseteq \mathbb{R}^n $ as $\mathcal{B}(J,S) \triangleq  \{x\in \mathbb{R}^n| J(x) \leq \rho \}$, where $\rho = \min _{x \in \partial S} J(x)$. For $a \in \mathbb{R}$, denote by $\lceil a \rceil$ the smallest integer larger than or equal to $a$.

\subsection{Optimal control of PWA systems}
We consider discrete-time PWA systems of the form
\begin{equation}\label{system}
	x_{t+1}=f_{\mathrm{PWA}}(x_t,u_t) =A_{i} x_t+B_{i} u_t+f_{i} \quad \text { if } \left[\begin{array}{l}
		x_t \\
		u_t
	\end{array}\right] \in \mathcal{C}_i,
\end{equation}
with input and state constraints $x \in X,\;u \in U$\footnote{All the results
	of this paper also apply to the case when there is a coupled
	constraint:$[x^T\;\;u^T]^T \in D$ with $D$ a polyhedron in $\mathcal{X} \times \mathcal{U}$, by letting $X$ be the projection of $D$ in $\mathcal{X}$ and by letting $U$ be a time-varying set depending on $x$.}. In \eqref{system}, $\{\mathcal{C}_{i}\}^s_{i=1}$ is a polyhedral partition of the state-input space $\mathcal{X} \times \mathcal{U}$. The matrices $A_i,B_i$ and the vectors $f_i$ define the affine dynamics in the regions $\mathcal{C}_i$. Including the offset vector $f_i$ allows for the representation of an affine transformation instead of just a linear one in each region of the state space. We want to design a control policy $u_t = \pi(x_t),\; t=0,1,\dots$ with $\pi: \mathcal{X} \rightarrow \mathcal{U}$ to simultaneously satisfy the constraints and minimize the infinite-horizon cost
\begin{equation}\label{cost}
	J\left( {{x_0},{\mathbf{u}}} \right) = J_\pi(x_0) =\sum\limits_{t = 0}^\infty  {l\left( {{x_t},\pi\left(x_{t}\right)} \right)},
\end{equation}
where ${\mathbf{u}} = \{u_t\}_{t=0}^{\infty} = \{\pi(x_t)\}_{t=0}^{\infty}$ stands for the infinite-horizon input sequence and the function $l : \mathbb{R}^{n_{x}} \times \mathbb{R}^{n_{u}} \rightarrow \mathbb{R}_{\geq 0}$ is the stage cost. %In this paper, we assume that the following conditions are always satisfied. 
Throughout the paper, we assume	that the dynamics, constraints, and stage costs satisfy the following assumption.
\begin{assumption}\label{A1}
	\begin{itemize}
		\item \emph{Dynamics:} The function $f_{\mathrm{PWA}}\left(\cdot, \cdot\right) :  \mathcal{X} \times \mathcal{U} \rightarrow \mathcal{X}$ is a continuous PWA function, with $\mathcal{X} \subseteq  \mathbb{R}^{n_x}$ and  $\mathcal{U} \subseteq \mathbb{R}^{n_u}$. %Here, $n_x$ and $n_u$ are the dimensions of the state and input vectors, respectively. 
		Besides, $f_{\mathrm{PWA}}\left(0, 0\right) =0$.
		
		\item \emph{Constraints:} The state constraint set $X = \bigcup_{i=1}^{r_0} X^{(i)} $ is a UoP, where $X^{(i)}$ is a polyhedron for each $i=1,\dots,r_0$ and $r_0$ is the number of polyhedra. The input constraint set $U$ is a polyhedron. Moreover, $X \times U  \subset  \bigcup_{i=1}^{s}{\mathcal{C}}_{i} = \mathcal{X} \times \mathcal{U}$.
		
		\item \emph{Stage cost:} The stage cost is based on the 1-/$\infty$-norm: $l\left( {{x_t},{u_t}} \right) = \|Q x_t\|_{q_1}+\|R u_t\|_{q_2}$, where $q_1,q_2 \in\{1, \infty\}$ and $Q, R$ have full column rank.  
	\end{itemize}
\end{assumption}

Similar assumptions can be found in other papers on control of PWA systems, e.g., \cite{baoti2006constrained,xu2016model}. In most literature \cite{baoti2006constrained,xu2016model,lazar2006stabilizing}, it is usually assumed that $X$ is a polyhedron, while we generalize it to a UoP. It should be mentioned that many practical nonlinear and non-convex state constraints, such as collision avoidance encountered in robotics, can be modeled or outer-approximated by constraint sets that are UoPs \cite{thirugnanam2022safety}. 

%The form of systems in \eqref{system} includes linear time-invariant systems by letting $s=1$. Besides, it also includes the case when the system is represented by a NN with PWA activation functions \cite{montufar2014number}, such as rectifier linear units (ReLU) and Max-Out functions.

By combining \eqref{system} and \eqref{cost}, our control objective is to solve the constrained infinite-horizon optimal control problem
%\begin{align}\label{problem1}
%		{J^*}\left( {{x_0}} \right) = &\mathop {\min }\limits_{\mathbf{u}} J\left( {{x_0},{\mathbf{u}}} \right) \nonumber \\
%		{\text{s}}{\text{.t}}{\text{.}}\;&{x_{t + 1}} = {f_{{\text{PWA}}}}({x_t},{u_t}),\;t = 0,1,... \nonumber \\
%	&x_t \in X,\;u_t \in U,\;t = 0,1,... 
%\end{align}
%or equivalently, solving the problem 
\begin{align}\label{problem2}
	{J^*}\left( {{x_0}} \right) = \mathop {\min }\limits_{\pi,\mathbf{u},\mathbf{x}}& J_\pi(x_0) \nonumber \\
	{\text{s}}{\text{.t}}{\text{.}}\;&{x_{t + 1}} = {f_{{\text{PWA}}}}({x_t},{u_t}),\;u_{t}=\pi\left(x_{t}\right), \nonumber \\
	&x_t \in X,\;u_t \in U,\;t = 0,1,... ,
\end{align}
%where ${\mathbf{u}}^*  =\left\{u^{*}_t\right\}_{t=0}^{\infty} $ is the optimizer of \eqref{problem1} (may not be unique), and
where ${\mathbf{x}} = \{x_t\}_{t=0}^{\infty}$, and $J^*(\cdot): \mathcal{X}  \to [0,\infty] $ is the optimal value function. An optimal policy, denoted by $\pi^{*}(\cdot)$, minimizes $J_{\pi}\left(x_{0}\right)$ subject to the constraints in \eqref{problem2} for any initial state $x_0$ that makes \eqref{problem2} feasible. In this paper, we denote by $\bar X$ the set of feasible initial states $x_0$ that make $J^{*}(x_0)$ finite. In \cite{borrelli2017predictive}, $\bar X$ is called the maximal stabilizable set for \eqref{system}. 

\begin{assumption}\label{A2}
	The set $\bar X$ is non-empty. Furthermore, for any $x_0 \in \bar X$, there exists a policy $\pi(\cdot)$ such that the system \eqref{system} with $u_t = \pi(x_t)$, starting from $x_0$, will reach the origin in a finite number of time steps.
\end{assumption}
Assumption \ref{A2} is a standard stabilizability assumption for discrete-time systems. Similar assumptions for PWA systems can be found, e.g., in \cite{lazar2006stabilizing,baoti2006constrained}.

For any $x \in \mathcal{X}$, according to Bellman’s Principle of Optimality \cite{bertsekas2019reinforcement}, the value function $J^*(\cdot)$ 
and the optimal policy $\pi^{*}(\cdot)$ satisfy the following equations:
\begin{align}\label{bellman}
	&J^{*}(x)=\Gamma J^{*}(x) \triangleq  \min _{u \in U}l(x, u)+J^{*}(f_\mathrm{PWA}(x, u)), \nonumber\\
	&\pi^{*}(x) \in \underset{u \in U}{\arg \min }\;l(x, u)+J^{*}(f_\mathrm{PWA}(x, u)),
\end{align}
where $\Gamma$ is called the Bellman operator \cite{bertsekas2019reinforcement}. In \eqref{bellman}, the domain of $J^{*}(\cdot)$ is the whole state space $\mathcal{X}$, which means that the value of $J^{*}(\cdot)$ goes to infinity outside $\bar X$. In general, the equation for $J^{*}(\cdot)$ in \eqref{bellman} may have multiple solutions. %For example, if $J^{*}(\cdot)$ is a solution we can conclude that $J^{*}(\cdot)$ plus an arbitrary constant is also a solution. 
Nevertheless, it follows from \cite[Proposition 1]{bertsekas2015value} that $J^{*}(\cdot)$ can be the unique solution that satisfies $J^{*}(0) = 0$ under Assumption \ref{A2}.

\subsection{Exact value iteration}
Solving the Bellman equations is in general computationally prohibitive for nonlinear systems. Usually, an MPC problem with a finite horizon is solved online to approximate the infinite-horizon optimal policy. However, computational complexity also remains a hurdle in the application of MPC to PWA systems. For the PWA system, the equations in \eqref{bellman}, on the other hand, can be solved by using an exact value iteration (VI) method \cite{baoti2006constrained}, which solves multiple multi-parametric linear programs (mp-LPs). To motivate our ADP methods, we summarize the exact VI method and discuss its limitations in this section. The exact VI algorithm starts from an initial value function $J_{0}(\cdot)$ that is either zero in $\mathcal{X}$ (case 1) or a control Lyapunov function defined on a subset of $\mathcal{X}$ (case 2). In case 2, the following assumption should be satisfied.

\begin{assumption}\label{A3}
	A continuous and PWA control Lyapunov function $J_\mathrm{CL}(\cdot): X_\mathrm{CI} \to \mathbb{R}_{\geq 0}$ in a polyhedral control-invariant set $X_\mathrm{CI}$ is available. In other words, $\min_ {u\in U,f_\mathrm{PWA}(x,u) \in X_\mathrm{CI}}\; l(x,u)+J_\mathrm{CL}(f_\mathrm{PWA}(x,u)) - J_\mathrm{CL}(x) \leq 0,\; \forall x \in X_\mathrm{CI}$.
\end{assumption}

Assumption \ref{A3} frequently appears in the stability analysis of MPC \cite{borrelli2017predictive}, where $J_\mathrm{CL}(\cdot)$ is chosen as the terminal cost and $X_\mathrm{CI}$ is specified as the terminal constraint. To satisfy Assumption \ref{A3}, it is sufficient to compute a stabilizing piecewise linear feedback law on $X_\mathrm{CI}$, and then $J_\mathrm{CL}(\cdot)$ can be computed by solving some nonlinear inequalities that contain 1-/$\infty$-norm of some linear functions \cite{lazar2006stabilizing}. 

With the initialization $X_{0}=X $ and $J_{0}(x)=0,\; \forall x \in X_{0}$ (case 1), or $X_{0}=X_\mathrm{CI} $ and $J_{0}(x)=J_\mathrm{CL}(x),\; \forall x \in X_{0}$ (case 2), the exact VI method iterates as follows:
\begin{align}\label{DP}
	J_{k}(x)=\min_{u \in U, f_{\mathrm{PWA}}(x, u) \in X_{k-1} } l(x, u)+J_{k-1}(f_{\mathrm{PWA}}(x, u))
\end{align}
for $k=1,2,..$. Here, $X_k = \mathrm{Pre}(X_{k-1}) \cap X_{0}$, where $\operatorname{Pre}(\mathcal{S})=\left\{x \in \mathcal{X} | \exists u \in U \text { s.t. } f_{\mathrm{PWA}}(x, u) \in \mathcal{S}\right\}$ is the backward-reachable set to a set $S$. Even for PWA systems with polytopic state constraints, the backward-reachable set to a polyhedral set can be a non-convex UoP because of the nonlinear dynamics \eqref{system}, which means that $X_k, k=1,2,\dots$ can be non-convex UoPs \cite{borrelli2005dynamic}.

%The exact VI algorithm in \eqref{DP} includes the procedure of computing the maximal control-invariant set (case 1) or the maximal stabilizable set $\bar{X}$ (case 2) for system \eqref{system}. More specifically, $X_{k}$ is the $k$-step controllable set \cite[Definition 2.9]{kerrigan2000invariant} for the system \eqref{system}. In case 1, it shrinks and converges to the maximal control-invariant set with respect to the Hausdorff distance \cite{kerrigan2000invariant,bertsekas1972infinite} as $k$ goes to infinity. If $\bar{X}$ is compact, the maximal control-invariant set will be equal to $\bar{X}$ \cite{borrelli2017predictive}. In case 2, $X_{k}$ further becomes the $k$-step stabilizable set \cite{borrelli2017predictive}, so it converges to $\bar{X}$ as $k$ goes to infinity \cite{blanchini1999set}. The computation of $X_{k}$ can be separated into three steps: (i) computing $\operatorname{Pre}\left(X_{k-1}\right)$; (ii) computing $\operatorname{Pre}\left(X_{k-1}\right) \cap X_{0}$ and (iii) removing redundant inequalities. These three steps can be efficiently implemented for PWA systems with linear constraints \cite{kerrigan2000invariant}, based on polyhedral operations.

The resulting $J^*(\cdot)$ and $\pi^*(\cdot)$ are both PWA functions sharing the same polyhedral partition of the feasible region $\bar X$ \cite{baoti2006constrained}.  However, the complexity (i.e., the number of polyhedral regions or affine functions) of $J^*(\cdot)$ and $\pi^*(\cdot)$ is exponential in both the dimension of the system and the number of constraints in \eqref{problem2} \cite{borrelli2003efficient}, so that storing the affine functions and regions of $J^*(\cdot)$ and $\pi^*(\cdot)$ needs a huge amount of memory. Secondly, the number of mp-LPs that need to be solved per iteration also grows exponentially with the problem dimension. Moreover, the online implementation of \cite{baoti2006constrained} needs to search which polyhedron the measured state belongs to. For high-dimensional systems, these regions may have complex representations. The growing complexities of the explicit controller's structure, and the offline and online computations limit the
applicability of exact VI to small-scale systems \cite{sadraddini2019sampling}. 

Based on these observations, it is thus necessary to simplify both the procedure of solving the Bellman equation and the control policy, by using some approximation methods.

However, the VI formulation in \eqref{DP} is not suitable for approximation because the probably non-convex constraint $f_{\mathrm{PWA}}(x, u) \in X_{k-1}$ leads to too complex optimization problems when using sample-based approaches.

\section{Value Iteration with Penalty Functions}
To deal with this issue, we consider soft state constraints by defining a penalty function $P(\cdot,\cdot)$. Suppose that each $X^{(i)}$ of the UoP constraint set $X=\bigcup_{i=1}^{r_0} X^{(i)}$ has the half-space representation $X^{(i)}\triangleq\{x\in \mathbb{R}^{n_x}| E_X^{(i)} x \leq g_X^{(i)}\}$, where $E_X^{(i)} \in \mathbb{R}^{m_x^{(i)} \times n_x}$, $g_X^{(i)} \in \mathbb{R}^{m_x^{(i)}}$, and $m_x^{(i)}$ is the number of rows of $E_X^{(i)}$. We design the penalty function $P(\cdot,\cdot)$ as the following min-max forms:
\begin{align}\label{penalty_define}
	P(x,X) &=  p \min_i \max \left\{0, (E^{(i)}_X)_{1\cdot}x  - (g^{(i)}_X)_{1}, \dots, \right.\nonumber\\
	&\quad \left.(E^{(i)}_X)_{m^{(i)}_x\cdot}x - (g^{(i)}_X)_{m^{(i)}_x} \right\}\; \text{or} \nonumber\\
P(x,X) &=  p \min_i \sum_{j=1}^{m^{(i)}_x} \max \left\{0, (E^{(i)}_X)_{j\cdot}x  - (g^{(i)}_X)_{j}\right\},
\end{align}
where  $(g^{(i)}_X)_{j}$ is the $j$th element of the vector $g^{(i)}_X$, $m^{(i)}_x$ is the number of rows of $E^{(i)}_X$, and $p>0$ is the constraint violation penalty weight. When $X$ reduces to a polyhedron, i.e., $r_0 =1$, the minimum operator in \eqref{penalty_define} will be removed.% and $P(\cdot,\cdot) $ is actually an exact penalty function that is often used in exact penalty methods \cite{bertsekas1997nonlinear}. 

An important property of $P(\cdot,\cdot) $ is that $P(\cdot,\cdot) $ is a PWA function w.r.t. its first argument. This means that adding such a penalty function into the cost function in \eqref{problem2} will not change the PWA properties of the optimal value function and the optimal policy. Note that we avoid barrier functions such as the logarithmic barrier function because they can go to infinity in any compact constraint sets and will deprive the value function of the PWA property. Besides, we do not penalize the input constraint violation, because the input constraints are single polyhedral constraints that can be readily handled.

In case 2 of \eqref{DP}, in addition to enforcing a soft penalty for $X$, we need to reconstruct the initial value function since $J_{\mathrm{CL}}(\cdot)$ is undefined outside $X_{\mathrm{CI}}$. To achieve this, we need to penalize $J_{{\text{CL}}}\left( x \right)$ for $x$ outside $X_{\mathrm{CI}}$ by finite values:
\begin{equation}\label{soft2}
	J_0^{ {\text{soft}}}(x) = \left\{ \begin{gathered}
		{J_{{\text{CL}}}}\left( x \right),\;\;\;\;\;\;\;\;\;\;\;\;\;\;\;\;\;\;\;\;x \in X_{\mathrm{CI}},  \hfill \\
		{J_{{\text{CL}}}}\left( {\bar z} \right) + P(x,X_{\mathrm{CI}}),\;x \notin X_{\mathrm{CI}}, \hfill \\ 
	\end{gathered}  \right.
\end{equation}
where $\bar z(\cdot)$ is one of the optimizers of the following mp-LP: 
\begin{equation}\label{mplp}
	\bar z(x) \in \arg\min_{z \in X_{\mathrm{CI}}}||z-x||_\infty.
\end{equation}  

In \eqref{soft2}, $J_0^{ {\text{soft}}}(\cdot)$ is continuous on $\mathcal{X}$, which will be proven in Theorem \ref{theorem1}. Based on the defined penalty function, a VI algorithm with penalty is developed as follows

\begin{algorithm}
	\caption{Value iteration with penalty}
	\label{alg:B}
	\begin{algorithmic}[1]
		
		\Statex \textbf{Output:} A value function $J^{\mathrm{soft}}_{k-1}(\cdot):\mathcal{X} \to \mathbb{R}_{\geq 0}$.
		
		\State Initialize the value function (option (a)) $J^{\mathrm{soft}}_{0}(x) \leftarrow 0,\; \forall x \in \mathcal{X}$, or (option (b)) by \eqref{soft2}. %这里初始0 或者Px都可

		\For {$k = 1,2,\dots$} 
		
		\State the value iteration $J_{k}^{\mathrm{soft}}(x) \leftarrow \Gamma_{\mathrm{p,\alpha}} J_{k-1}^{\mathrm{soft}}(x),\; \forall x \in \mathcal{X}$, where $\alpha = 1$ if option 1 is chosen, or $\alpha = 2$ if option 2 is chosen, and $\Gamma_{\mathrm{p,\alpha}}$ is defined in \eqref{algo3} and \eqref{modify}.
		
		\State If $J_{k}^{\mathrm{soft}}(x) = J_{k-1}^{\mathrm{soft}}(x),\;\forall x \in \mathcal{X}$, \textbf{break}.
		\EndFor
	\end{algorithmic}
\end{algorithm} 

In Algorithm \ref{alg:B}, we consider two options for the VI, in which we define two Bellman operators for $J: \mathcal{X} \to \mathbb{R}$. The first one used in option 1 is
\begin{equation}\label{algo3}
\Gamma_{\mathrm{p,1}} J(x)\triangleq \min _{u \in U} \;l_\mathrm{p}(x, u)+J\left(f_{\mathrm{PWA}}(x, u)\right),  x \in \mathcal{X},
\end{equation}
where $l_\mathrm{p}(x, u) = l(x,u)+P(x,X)$. The second one used in option 2 is
\begin{align}\label{modify}
	\Gamma_{\mathrm{p,2}} J(x)\!\triangleq\! \min _{u \in U} \;l(x, u)\!+\!J\left(f_{\mathrm{PWA}}(x, u)\right)&\!+\! P_{k-1}\left(f_{\mathrm{PWA}}(x, u)\right), \nonumber\\
	& x \in \mathcal{X},
\end{align}
where $P_0(x) = 0, \forall x \in \mathcal{X}$ and $P_k(x) = P(x, X), \forall x \in \mathcal{X}$ and $\forall k>0$. Since the state constraints are removed, the working region of VI is the whole state space $\mathcal{X}$. Besides, we also consider two options (options (a) and (b)) for the initialization of the value function. The combinations of above options result in four different options: options 1(a), 1(b), 2(a), and 2(b). Note that adding a penalty into the stage cost, which is applied in option 1 of Algorithm \ref{alg:B}, is common in existing constrained ADP methods \cite{xu2022adaptive}. In comparison, in option 2 of the algorithm, we propose a novel scheme in which we add a penalty into the cost-to-go $J_{k-1}^{\mathrm{soft}}\left(f_{\mathrm{PWA}}(x, u)\right)$. In the following theorem, we will analyze the PWA property and continuity of each $J_{k}^{\mathrm{soft}}(\cdot)$ as well as the convergence of the sequence $\{J_{k}^{\mathrm{soft}}(\cdot)\}^\infty_{k=0}$ to a fixed optimal value function in all options. 
	
\begin{theorem}\label{theorem1}
	Considering Algorithm \ref{alg:B}, if Assumptions \ref{A1}-\ref{A2} hold in option (a) and Assumptions \ref{A1}-\ref{A3} hold in option (b), each $J_{k}^{\mathrm{soft}}(\cdot),\; k<\infty$ is a continuous PWA function on $\mathcal{X}$ and the value function sequence $\{J_k^{\mathrm{soft }}(x)\}^\infty_{k=0}$ converges point-wise to
	\begin{align*}\label{batch0}
		{J^{\mathrm{soft}*}}\left( {{x}} \right) = \mathop {\min }\limits_{\{u_i,x_i\}^{\infty}_{i=0}}& \sum_{i=0}^{\infty} l_\mathrm{p}\left(x_i, u_i\right)+ P(x_i,X)  \nonumber \\
		\mathrm{s.t.}\;\;\;\;&{x_{i + 1}} = {f_{{\text{PWA}}}}({x_i},{u_i}),\;u_i \in U,\;i = 0,1,... \nonumber \\
		&x_0 =x 
	\end{align*}
	for all $x \in X$. 
\end{theorem}

Option 2, incorporating penalties into the cost-to-go, yields the equivalent value function $J_k^{\text {soft }}(\cdot)$ as option 1, which adds penalties to stage costs. Both options can alleviate the violation of state constraints by adding penalties to the overall infinite-horizon cost function. Detailed comparison between options 1 and 2 is given in Section IV. C. 

\textcolor{blue}{After the optimal value function $J_k^{\text {soft }}$ is obtained by Algorithm 1, the control policy is implicitly determined by the solution of the optimization problem \eqref{bellman} with $J^*$ replaced by $J_k^{\text {soft }}$.} 

\begin{remark}
	If the PWA system reduces to a linear time-invariant system and $X$ is a polyhedron, it is possible to make $J_k^{\mathrm{soft }}(x) = J_k(x), \forall x\in X_k$, by choosing a large enough $p$, according to the exact penalty theorem \cite[Proposition 5.4.5]{bertsekas1997nonlinear}.
\end{remark}

\section{Constrained ADP algorithm}
\subsection{Algorithm design}
Continuity of the value functions, established in Theorem \ref{theorem1}, is desired since it enables a universal approximation capability \cite{hanin2019universal}. With Algorithm \ref{alg:B}, a tractable ADP approach can be developed to approximate each $J_k^{\mathrm{soft}}(\cdot)$. In particular, a function approximator (critic) $\hat{J}_k\left(\cdot, \theta_k\right)$, which is parameterized by $\theta_k$, is constructed to replace $J_k^{\mathrm{soft}}(\cdot)$. In each iteration $k$, a set $X_\mathrm{s}= \{x^{(i)}\}^{N_x}_{i=1}$ of state samples is collected from a compact region of interest $\Omega_k \subseteq \mathcal{X}$, according to some strategies such as sampling from a uniform grid and random sampling \cite{busoniu2017reinforcement}. Here, $N_x$ is the number of samples. The update of the parameter $\theta_k$ minimizes $\sum_{i=1}^{N_x}[\Gamma_{\mathrm{p}, \alpha} \hat J_{k-1} (x,{\theta _{k-1}} )|_{x=x^{(i)}}- {\hat J_k}( {x^{(i)},{\theta _k}} )]^2$. The iterative procedure stops when the difference between $\theta_k$ and $\theta_{k-1}$ is small enough. The detailed procedure is given in Algorithm \ref{algo1}.

\begin{algorithm}
	\caption{Constrained approximate value iteration}
	\label{algo1}
	\begin{algorithmic}[1]
		\Statex \textbf{Output:} A value function $\hat{J}_{k-1}\left(\cdot, \theta_{k}\right)$
		\State Option (a): Initialize the value function $\hat{J}_0(\cdot, \theta_0) \leftarrow 0, \forall x\in \Omega_0$, where $\Omega_0 = \mathcal{X}$ in option 1(a) or $\Omega_0 = X$ in option 2(a).
		
		\Statex Option (b): Initialize the value function $\hat{J}_0(\cdot, \theta_0)$ by
		\begin{equation}\label{evaluation}
			{\theta_0} \leftarrow \mathop {\arg \min }\limits_\theta  \sum\limits_{x^{(i)} \in X_\mathrm{s} \cap  \Omega_0} {{{ \rho_v(x^{(i)})( { J_0^\mathrm{soft}(x^{(i)})- {{\hat J}_0}({x^{(i)}},{\theta })} )}^2}},
		\end{equation}
		where $J_0^\mathrm{soft}(\cdot)$ is from \eqref{soft2}, and $\Omega_0 = \mathcal{X}$ in option 1(b) or $\Omega_0 = X_\mathrm{CI}$ in option 2(b).
		\For {$k = 1,2,\dots$} 
		\If{option 1 is chosen,} let $\Omega_k \leftarrow \mathcal{X}$ and $\alpha \leftarrow 1$.\EndIf
		\If{option 2 is chosen,} let $X_k \leftarrow \operatorname{Pre}\left(X_{k-1}\right) \cap X $, $\Omega_k \leftarrow X_k $, and $\alpha \leftarrow 2$.\EndIf
		\State Obtain the target value $v^{(i)}_k$ by 
		\begin{equation}\label{target}
			v^{(i)}_k \leftarrow \Gamma_{\mathrm{p,\alpha}} \hat{J}_{k-1}(x,\theta_{k-1})|_{x=x^{(i)}}, \forall x^{(i)} \in X_\mathrm{s} \cap \Omega_k
		\end{equation}
		% 		\If{$f_\mathrm{PWA}(x^{(i)}, u^{(i)}_k) \in X_{k-1}$}
		% 		\State $\mathcal{S}   \leftarrow  \mathcal{S}  \cup  \{x^{(i)}\}$
		% 		\EndIf
		\State Find $\theta_k$ such that 
\begin{equation}\label{update}
	{\theta _k} \!\leftarrow\! \mathop {\arg \min }\limits_\theta  \!\sum\limits_{x^{(i)}\! \in \!X_\mathrm{s} \!\cap \Omega_k  }\! {{{ \rho_v(x^{(i)})( {v_k^{(i)}\! - \!{{\hat J}_k}({x^{(i)}},{\theta })} )}\!^2}}. 
\end{equation}	
		\If{$|\hat{J}_{k}(x,\theta_{k}) - \hat{J}_{k-1}(x,\theta_{k-1})| \leq \epsilon(x),\;\forall x \in \Omega_k \cap \Omega_{k-1}$,} \textbf{break}. \EndIf
		\EndFor
	\end{algorithmic}
\end{algorithm} 

In \eqref{update} of step 8 of Algorithm \ref{algo1}, $\rho_v(\cdot): \mathcal{X} \to \mathbb{R}_{>0}$ is the state relevance weighting function. In step 9, $\epsilon(\cdot): \mathcal{X} \to \mathbb{R}_{\geq0}$ is a tolerance function, determining whether $\hat{J}_{k-1}\left(\cdot, \theta_{k-1}\right)$ is satisfactory. Both of $\rho_v(\cdot)$ and $\epsilon(\cdot)$ will be designed later in Section V.A. Besides, in practice one would also need a limit on the maximum number of iterations.

%It is noticed from Theorem \ref{theorem1} that $J_k^{\mathrm{soft }}(x)$ could reach infinity for some $x \in  \mathcal{X}/X_k$ unless the following assumption on the global stabilizability with only the input constraint is made.
%\begin{assumption}\label{A4}
%	$J^{\mathrm {soft }*}(x) < \infty, \forall x\in \mathcal{X}$.
%\end{assumption}
%
%Assumption \ref{A4} ensures that $J_k^{\mathrm{soft }}(x) < \infty, \forall x\in \mathcal{X}$ and $\forall k \in \{0,1,...\}$, and thus makes the approximation of $J_k^{\mathrm{soft }}(\cdot)$ on $\mathcal{X}$ reasonable. Therefore, Assumption \ref{A4} is supposed to hold when choosing option 1 of Algorithm 2. In option 2, Assumption \ref{A4} is not required because $J_k^{\mathrm{soft }}(\cdot)$ is finite on $X_k$.

With $\hat{J}_{k-1}\left(\cdot, \theta_{k-1}\right)$ available, a sub-optimal control policy $\hat{\pi}^\mathrm{im}(x)$ can be implicitly determined by
\begin{equation}\label{suboptimal}
	\hat{\pi}^\mathrm{im}(x) \in \arg\min _{u \in U}\;l(x, u)+\hat{J}_{k-1}(f_\mathrm{PWA}(x, u),\theta _{k-1}),\; \forall x \in \Omega_{k}
\end{equation}
%Besides, if option 2(b) of Algorithm \ref{algo1} is implemented, since each $\Omega_{k},\; k=0,1,...$ is control-invariant \cite[page 195]{borrelli2017predictive}, we can get another implicit policy 
%\begin{align}\label{suboptimal2}
%	\hat{\pi}^\mathrm{im,2}(x) \in \arg&\min _{
%			u \in U} \;l(x, u)+\hat{J}_{k-1}(f_\mathrm{PWA}(x, u),\theta _{k-1})\nonumber\\
%	 &\mathrm{s.t.} \; {f_{{\text{PWA}}}}(x,u) \in \Omega_{k-1},\;\forall x \in \Omega_{k}
%\end{align}
%which can ensure the closed-loop system to be recursively feasible in $\Omega_{k}$. 

In step 1 of Algorithm \ref{algo1}, the function approximator is initialized by regressing $J_0^\mathrm{soft}(\cdot)$, provided that the explicit form of $J_{\mathrm{CL}}(\cdot)$ is known. If the explicit form of $J_{\mathrm{CL}}(\cdot)$ is not available but a stabilizing and safe piecewise linear feedback law $\pi_{\mathrm{PWL}} (\cdot) : \mathcal{X}\to \mathcal{U}$ on $X_\mathrm{CI}$ is known, we can initialize $\hat{J}_0\left(\cdot, \theta_0\right)$ as an approximation of $J_{\pi_{\mathrm{PWL}}} (\cdot) $, which is also a control Lyapunov function, by doing a policy evaluation \cite{busoniu2017reinforcement}.

% ADP algorithms are valid only if the entire state trajectory starting from the initial state remains within $\Omega$ \cite{heydari2018stability}. It can be the case when $f_{\mathrm{PWA}}\left(x, u\right)$ is outside $\Omega$ for some $x \in \Omega$ and $u$ so that $\hat{J}_{k-1}\left(f_{\mathrm{PWA}}\left(x, u\right), \theta_k\right)$ is no longer a good approximation of $J_{k-1}^{\mathrm{soft}}\left(f_{\mathrm{PWA}}(x, u)\right)$. This can cause errors in the approximation of $\hat{J}_{k}$ in the next iteration. Such errors cannot be reduced by increasing the number of samples or the complexity of the approximator. Moreover, in our situation where there are state constraints, to implement the above normal ADP procedure, $\Omega$ must be larger than $X$ so that the value of $\hat{J}_{k}$ for those samples outside $X$ can be penalized by the steep rise of $l_{\mathrm{p}}(\cdot,\cdot)$. A large $\Omega$, on the other hand, imposes a burden on the approximation capability of the function approximator.

To carry out the iterative procedure in Algorithm 2 more efficiently, we need to use a proper function approximator at each iteration. Since each $J^\mathrm{soft}_k(\cdot)$ is a PWA function, it is preferable that the candidate approximator can also output a PWA function. Suitable choices are thereby NNs with (leaky) rectifier linear units (ReLUs) as activation functions, difference of two max-affine functions, Min-Max NNs \cite{lohmiller2021deep}, and so on. Detailed descriptions of these function approximators are given in the Appendix \ref{MILP formulation}.

\emph{Computation of $X_k$ under the UoP state constraint:} In option 2, one is required to compute the reachable sets $X_k$. The following lemma states that we are able to obtain $X_k$ by performing polyhedral operations even if the state constraint is UoP. The proof is given in Appendix \ref{lemma1proof}.
\begin{lemma}\label{computex}
	For the PWA system \eqref{system} with a convex polyhedral input constraint $u \in U$ and a UoP state constraint $x \in X =\bigcup_{i=1}^{r_0} X^{(i)} $, the set iterates $X_k = \mathrm{Pre}(X_{k-1}) \cap X, \;k=1,2,\dots$ make each $X_k$ be a UoP. Then, if $X_{k-1}$ is in the form of $X_{k-1} = \bigcup_{i=1}^{r_{k-1}} X^{(i)}_{k-1}$, where each $X^{(i)}_{k-1}$ is a polyhedron, then $X_k$ can be computed by 
	\begin{equation}\label{eqcomputex}
		X_k = \bigcup_{j=1}^{r_0} \bigcup_{i=1}^{r_{k-1}} \mathrm{Pre}(X^{(i)}_{k-1})\cap X^{(j)}
	\end{equation}
\end{lemma}

\emph{Mixed-integer formulations of problems \eqref{target} and \eqref{suboptimal}:} Problems \eqref{target} and \eqref{suboptimal} have similar forms. They can be transformed into MILP problems since both the PWA system and PWA function approximators are MILP representable, which is shown in Appendix \ref{MILP formulation}. Here, we say a function $J$ is MILP representable if $J$ can be represented by a set of mixed-integer linear equations and inequalities containing additional variables. We note this set by $\mathrm{gr}_\mathrm{MILP}(J)$. As a result, we obtain mixed-integer formulations of problems \eqref{target} and \eqref{suboptimal}.
%\begin{definition}[\cite{schwan2022stability}]
%	A function $h: \mathbb{R}^n \to \mathbb{R}^m$ is MILP representable if there exists a polyhedral set $\mathcal{P} \subseteq \mathbb{R}^n \times \mathbb{R}^m \times \mathbb{R}^c \times \mathbb{R}^b$ such that $(x,u) \in \mathrm{gr}(h)$ iff there exist $z \in \mathbb{R}^c$ and $\beta \in \{0,1\}^b$ such that $(x,u,z,\beta) \in \mathcal{P}$.
%\end{definition}
Precisely, if, e.g., the first type of the penalty function in \eqref{penalty_define} and the infinity norm of the stage cost are chosen, \eqref{target} in option 1 can be equivalently written as the following MILP:
\begin{align*}
	&\min _{\substack{u,\varepsilon^{(x)},\varepsilon^{(u)},\\ \varepsilon^{(J)} ,\varepsilon^{(P)},x^+}}\; \varepsilon^{(x)}+ \varepsilon^{(u)} + \varepsilon^{(J)} +\varepsilon^{(P)} \nonumber\\ 
	&\mathrm{s.t.}  -1_{n_x} \varepsilon^{(x)} \leq \pm Q x,\;-1_{n_u} \varepsilon^{(u)} \leq \pm R u , \; u \in U,\nonumber\\
	&  \quad  (x^+,\varepsilon^{(J)}) \in \mathrm{gr}_\mathrm{MILP}(\hat{J}_{k-1}), \;  (x,\varepsilon^{(P)}) \in \mathrm{gr}_\mathrm{MILP}(P(\cdot,X)), \nonumber\\
	&  \quad ([x^T\;u^T]^T,x^+) \in \mathrm{gr}_\mathrm{MILP}(f_\mathrm{PWA}).
\end{align*}
Similar transformations can be achieved for problem \eqref{suboptimal}, for option 2 of Algorithm \ref{algo1}, and for the other choice of $P(\cdot,\cdot)$ and $l(\cdot,\cdot)$.

MILP problems can be effectively solved by using the branch-and-bound approach \cite{wolsey1999integer}, which is a global optimization algorithm.

Different from \eqref{target}, Problem \eqref{suboptimal} is solved online. Even through it still belongs to an MILP problem, \eqref{suboptimal} can be solved more rapidly than a general hybrid MPC problem with a long horizon, because \eqref{suboptimal} in general results much fewer auxiliary and binary variables. That is one of main benefits of using ADP or RL.
\begin{remark}
When there are approximation errors, the convergence of $\hat{J}_{k}(\cdot,\theta_{k})$ to $J^{\text {soft* }}(\cdot)$ is in general not guaranteed because the infinite-horizon cost \eqref{problem2} is undiscounted and does not induce a contraction property for the Bellman operator. In general, one can add a discount factor to \eqref{problem2} to ensure the convergence of the value iteration under approximation errors, but this may come out the cost of weakening the stability \cite{postoyan2016stability}.
\end{remark}
\subsection{Approximating explicit policies}

Since two PWA functions $f_\mathrm{PWA}(\cdot,\cdot)$ and $\hat{J}_{k-1}(\cdot, \theta_{k - 1})$ are coupled in \eqref{suboptimal}, \eqref{suboptimal} may still have many auxiliary and binary variables, if, e.g., a multiple-layer (deep) NN is used. 
As \eqref{suboptimal} needs to be solved online, the advantage of low computational complexity brought by ADP is not obvious. To avoid solving complex MILP problems online, the policy $\hat{\pi}^\mathrm{im}(\cdot)$ can also be represented explicitly, in which case it usually needs to be approximated by a second function approximator (actor). The actor is also recommended to having a PWA form since the optimal control policy $\pi^{*}(\cdot)$ is PWA. 

%这里删去了第二个 im ex policy,因为加入下一状态的约束会导致稳定性证明中无法放缩
As the optimizer $\hat{\pi}^\mathrm{im}(\cdot)$ can be discontinuous and not unique, instead of using supervised learning methods to train the actor, we can directly construct a parameterized policy $\hat{\pi}^\mathrm{ex}(\cdot,\omega)$ with parameter $\omega$ and update $\omega$ to minimize the expectation of the objective function in \eqref{suboptimal} w.r.t. the sample distribution $d_s$ used in Algorithm \ref{algo1}. This results in the following policy optimization problem
\begin{align}\label{policy}
	 &\begin{aligned}
	\omega^* \in \arg\min _{
		\omega} \; \mathbb{E}_{x  \sim d_s}& [\rho_\pi(x)(l(x, \hat{\pi}^\mathrm{ex}(x,\omega)) \\
	&+\hat{J}_{k-1}(f_\mathrm{PWA}(x, \hat{\pi}^\mathrm{ex}(x,\omega)),\theta _{k-1}))]
\end{aligned}\nonumber\\
	&\quad \quad \quad \quad  \mathrm{s.t.} \; \mathbb{E}_{x  \sim d_s}[\hat{\pi}^\mathrm{ex}(x,\omega)] \in U,
\end{align}
where $\rho_\pi (\cdot): \mathcal{X} \to \mathbb{R}_{>0}$ is another state relevance weighting function to be specified later in Section V.A. Similar to the critic, we specify $\hat{\pi}^\mathrm{ex}(\cdot,\omega)$ as a PWA approximator.

To solve \eqref{policy}, the policy gradient method, combined with the Lagrangian relaxation approach \cite{tessler2018reward} or the augmented Lagrangian approach \cite{li2021augmented} for constraints handling, can be employed. 

The above procedures are conducted offline. Ideally, if there are no approximation errors on both the critics and the actor, and the penalty weight $p$ and the number of iterations are infinite, we have $X_\infty = \bar X$ and $\hat{\pi}^\mathrm{ex}(\cdot,\omega^*) = \hat{\pi}^\mathrm{im}(\cdot)$. Consequently, $\hat{\pi}^\mathrm{ex}(\cdot,\omega^*)$ will always make the system satisfy all the constraints for the initial condition $x_0 \in \bar X$. However, due to approximation errors and the finite penalty weight, the policy cannot always satisfy the state and input constraints. As the input constraints are usually hard constraints, in the online setting we project $\hat{\pi}^\mathrm{ex}(\cdot,\omega^*)$ onto $U$ when the current state $x_t$ is received. This results in a convex quadratic program:
\begin{equation}\label{project}
	\phi(u^\mathrm{ex}_t) = \arg\min _{u \in U} ||u-u^\mathrm{ex}_t||_2, 
\end{equation}
where $u^\mathrm{ex}_t = \hat{\pi}^{\mathrm{ex}}\left(x_t, \omega^*\right)$, and the function $\phi(\cdot) : \mathbb{R}^{n_u} \to \mathbb{R}^{n_u}$ maps the output of the actor to its projected value. Problem \eqref{project} can be treated as a parametric quadratic program with the parameter $u_t^{\mathrm{ex}}$. Therefore, the optimizer $\phi(\cdot)$ of \eqref{project} is unique, PWA \cite[Theorem 6.7]{borrelli2017predictive}, and also MILP representable \cite[Lemma 4]{schwan2022stability}. Meanwhile, \eqref{project} defines a projected policy $\hat{\pi}^\mathrm{ex}_\mathrm{proj}(\cdot) = \phi(\hat{\pi}^{\mathrm{ex}}\left(\cdot, \omega^*\right)) $ of $\hat{\pi}^{\mathrm{ex}}\left(\cdot, \omega^*\right)$.
\begin{remark}
The number of decision variables and the number of constraints in the convex QP problem \eqref{project} are not larger than those in the MILP problem \eqref{suboptimal}. It is known that convex QP problems are P problems while MILP problems are NP hard problems. Therefore, the consideration of the explicit policy $\hat{\pi}^{\mathrm{ex}}\left(x_t, \omega^*\right)$ further enhances the online computational efficiency compared to \eqref{suboptimal}.
\end{remark}

\subsection{Discussions on the ADP method}
\emph{Comparison between options 1 and 2 of Algorithm \ref{algo1}:} The main differences between options 1 and 2 are the training region $\Omega_{k}$ and the way each $\hat{J}_{k-1}(\cdot,\theta_{k-1})$ iterates. These differences lead to differences in the adaptability and efficiency of options 1 and 2:
\begin{itemize}
	%\item Option 1 requires Assumption \ref{A4} while option 2 does not.
	\item Compared to option 1, option 2 is more efficient in sampling and can result in a better approximation accuracy, because the working region $\Omega_{k}$ in option 2 is in general much smaller than $\mathcal{X}$. This is also the main advantage of adding penalties into the cost-to-go over adding penalties into the stage cost. We note that to implement option 1, one should choose a region of interest for sampling, and the region must be larger than $X$, so that the constraint violation can be penalized in the critic. However, the states that can be steered to the origin and have zero constraint violation are all contained in $\bar X$, which is much smaller than $\mathcal{X}$. Accordingly, in option 2, we concentrate on $X_{k}$, which converges to $\bar X$ as $k$ goes to infinity.
	\item On the other hand, option 2 needs to compute the $k$-step controllable set $\Omega_k$ ($X_k$) while option 1 does not. Therefore, for large-scale PWA systems, for which the exact computation of each $X_k$ is computationally very demanding, option 1 is preferable. %Moreover, in the worst case, i.e., when the system is in large scale and Assumption \ref{A4} is not satisfied, to implement option 2, we iteratively compute a sequence of zonotopes $\hat X_k = \{\bar{x}_k\}\oplus G_k \mathbb{P},\;k=1,2,... $, where $\bar{x}_k \in \mathbb{R}^{n_x}$, $G_k \in \mathbb{R}^{n_x \times n_x} $, and $\mathbb{P}=[-1,1]^{n_x}$. \cite[Section III]{sadraddini2019sampling} provides a way to make $\hat X_k$ an inner approximation of $\operatorname{Pre}\left(\hat{X}_{k-1}\right) \cap X$ with $\hat X_0 = X$, by solving an MILP. The complexity thus grows exponentially with the number of integers, which is $s+r_0$ rather than the system's dimension. 
\end{itemize}

\emph{Comparison to RL with a safety filter:} In the schemes of \cite{li2022robust,wabersich2021predictive}, RL policies are projected onto a safe set where both state and input constraints are considered. That design is motivated from the fact that the policies in \cite{li2022robust,wabersich2021predictive} are derived from standard RL algorithms that do not account for constraints. In comparison, our proposed ADP algorithms incorporate the state constraints into the cost function by adding penalty terms for violating the constraints. Besides, input constraints are regarded as hard constraints in the optimization problems \eqref{target}, \eqref{suboptimal}, and \eqref{project}. The optimal value function $J^{\mathrm{soft} *}(\cdot)$ with penalties should be less than or equal to the optimal value function $J^*(\cdot)$ of the original constrained optimal control problem \eqref{problem2}, as the optimal solution to \eqref{problem2} is consistently feasible for the unconstrained optimal control problem in Theorem 1. However, the potential over-optimality may result from minor violations of the state constraints. Therefore, in Section V we provide a tool to offline verify the state constraint satisfaction.

\emph{Remark:} We can strictly guarantee the state constraint satisfaction a priori by projecting the successor state onto a controlled-invariant set \cite{li2022robust}. This approach, however, will make the online computational complexity very large.

\section{Performance analysis and verification}
In this section, we will characterize the stability and safety of the closed-loop system with the policies $\hat{\pi}^\mathrm{im}(\cdot)$ and $\hat{\pi}^\mathrm{ex}_\mathrm{proj}(\cdot)$, and also the sub-optimality properties of these policies. Firstly, we provide general conditions under which stability and safety hold. These conditions can guide the parameter tuning of Algorithm \ref{algo1}. Then, we give sub-optimality guarantees, i.e., a bound on the mismatch between the infinite cost of the policies and real value functions. Finally, we develop verifiable stability and safety conditions. We say that a closed-loop system is safe in a set if its states and inputs satisfy the constraints for all trajectories starting from the set. 
\subsection{Stability and safety analysis}
First, we state a useful lemma that gives some properties of the value function $J_k^{\mathrm{soft}}(\cdot)$.
\begin{lemma}\label{lemma2}
	Consider Algorithm 1. Suppose that Assumptions \ref{A1}-\ref{A2} hold in option (a) and Assumptions \ref{A1}-\ref{A3} hold in option (b).
	
	(i) Then, there exists a positive constant $\gamma<\infty$ such that for all $k \geq 0$, $J_k^{\mathrm{soft}}(x) \leq \gamma l(x,0)$, $\forall x \in \bar{X}$;
	
	(ii) there exists a finite $\bar{k}>0$ such that for all $k\geq\bar{k}$, we have
	\begin{equation}\label{stable}
		J_k^{\mathrm{soft}}(x)-J_{k-1}^{\mathrm{soft}}(x) \leq \beta l(x,0),\;\forall x \in \bar{X} ,\; \mathrm{with}\; \beta \in (0,1).
	\end{equation}
\end{lemma}
The proof of Lemma \ref{lemma2} is in Appendix \ref{lemma2proof}. Then, we are ready to state the main result in this subsection. 

\begin{theorem}\label{stable2}
	Consider Algorithm \ref{algo1} and the proposed policies $\hat{\pi}^\mathrm{im}(\cdot)$ and $\hat{\pi}^\mathrm{ex}_\mathrm{proj}(\cdot)$. Let $\Omega$ be a compact subset of $X$. Suppose that Assumptions \ref{A1}-\ref{A2} hold in option (a) and Assumptions \ref{A1}-\ref{A3} hold in option (b). Consider the follow conditions:
	
	(C1): There exist a constant $\zeta \in (0,1)$ and a positive integer $k$ such that $
		\left| {{{{{\hat J}_{k - 1}}(x) - J_{k-1}^\mathrm{soft} (x)}} } \right| \leq  	\zeta {{J_{k-1}^\mathrm{soft} (x)}},\;\forall x \in \Omega$.
	
	(C2): There exist  a constant $e_\mathrm{p}>0$ and a positive integer $k$ such that $ \hat{J}_{\hat{\pi}^\mathrm{ex}_\mathrm{proj}} (x)- \hat{J}_{\hat{\pi}^\mathrm{im}} (x)
	 \leq e_\mathrm{p} l(x,0),\;\forall x \in \Omega$. Here, $\hat{J}_{\pi}(\cdot)$ is defined by $\hat{J}_{\pi} (x) \triangleq l(x,\pi(x)) + \hat{J}_{k-1}(f_\mathrm{PWA}(x, \pi(x)))$.
	
	As a result, we have:
	
	(i) If C1 holds with $k \geq \bar{k}$ and 
	\begin{equation}\label{zeta11}
		(1+\zeta)(1-\beta) > \max(2\zeta \gamma,\;1),
	\end{equation}
	where $\bar{k}$, $\beta $, and $\gamma$ come from Lemma \ref{lemma2}, the closed-loop system $x_{t+1} = f_\mathrm{PWA}(x_t,\hat{\pi}^\mathrm{im}(x_t)),\;t=0,1,...$ is asymptotically stable and safe in $\mathcal{B}(J^{\mathrm{soft}}_{k-1},\Omega) \cap \mathcal{B}(\hat{J}_{k-1},\Omega)$.
	
	(ii) If C1 and C2 hold with $k \geq \bar{k}$, and 
		\begin{equation}\label{zeta12}
		(1+\zeta)(1-\beta) > \max(2\zeta \gamma+e_\mathrm{p} ,\;1),
	\end{equation}
	the closed-loop system $x_{t+1} = f_\mathrm{PWA}(x_t,\hat{\pi}^\mathrm{ex}_\mathrm{proj}(x_t)),\;t=0,1,...$ is asymptotically stable and safe in $\mathcal{B}(J^{\mathrm{soft}}_{k-1},\Omega) \cap \mathcal{B}(\hat{J}_{k-1},\Omega)$.
\end{theorem}

\begin{remark}
	\textcolor{blue}{Although Theorem \ref{stable2} provides sufficient conditions for stability, some of them (e.g., C1 and C2) are difficult to verify. A method that can verify C1 and C2 in a probabilistic way is reported in \cite{hertneck2018learning}. On the other hand, Theorem \ref{stable2} suggests several ways to design the parameters and function approximators in Algorithm \ref{algo1}. For practical verification of stability and safety, in Section V.C, we move beyond using Conditions C1 and C2. Instead, we propose an offline verification framework based on solving MILP problems. As will be demonstrated in the case study, this framework effectively verifies safe and stable regions.}
\end{remark}

\begin{itemize}
	\item All results in Theorem \ref{stable2} require \eqref{stable} to hold. The left-hand side of \eqref{stable} is about the residual error of VI. It indicates that a suitable tolerance function $\epsilon(\cdot)$, which determines the stopping condition at step 9 of Algorithm \ref{algo1}, could be $\epsilon(x) = e_\mathrm{tole} l(x,0)$, for some $e_\mathrm{tole} \in (0,1)$. 
	
	\item Condition C1 limits the mismatch between ${\hat J}_{k - 1}(\cdot)$ and $J_{k-1}^\mathrm{soft} (\cdot)$, which further limits the approximation error $ \upsilon_i(x) \triangleq \hat{J}_i(x)-\Gamma_{\mathrm{p}, \alpha}\hat{J}_{i-1}(x),\;i=1,...,k$ of VI. To make $\zeta$ as small as possible, which helps to fulfill \eqref{zeta11} and \eqref{zeta12}, $\rho_v(\cdot)$ in \eqref{update} could be specified by  $\rho_v(x)= 1/l^2(x,0)$. To understand this, we consider the state trajectory $x_0,\;x_1,...,x_k$ that is generated from the closed-loop system $x_{t+1}=f_\mathrm{PWA}(x_t, \pi_{k-t}(x_t)),\;t=0,...,k-1$, where $\pi_i(\cdot)$ denotes the optimizer of $\Gamma_{\mathrm{p}, \alpha} J^\mathrm{soft}_{i-1}$. Then, we have
	\begin{align*}
		\hat{J}_k(x_0)-J^\mathrm{soft}_{k}(x_0)& \leq \hat{J}_{k-1}(x_1)-J^\mathrm{soft}_{k-1}(x_1)+ \upsilon_k(x_0)\nonumber\\
		&\leq \hat{J}_{0}(x_k)-J^\mathrm{soft}_{0}(x_k)+ \sum_{i=1}^{k} \upsilon_i(x_{k-i}).
	\end{align*}
	Suppose that $|\hat{J}_{0}(x_k)-J^\mathrm{soft}_{0}(x_k)| \leq e_\mathrm{v}  l(x_k,0)$ and $|\upsilon_i(x_{k-i})| \leq e_\mathrm{v}  l(x_{k-i},0),\;i=1,...,k$, where $e_\mathrm{v}>0$, we obtain 
	\begin{equation}\label{23}
		\hat{J}_k(x_0)\!-\!J^\mathrm{soft}_{k}(x_0) \leq e_\mathrm{v}  \sum_{i=0}^{k} l(x_i,0) \leq e_\mathrm{v} J_k^{\mathrm{soft }}(x_0).
	\end{equation}
Similar procedures can be applied to upper bound the value of $J^\mathrm{soft}_{k}(x_0)-\hat{J}_k(x_0)$. From \eqref{evaluation} and  \eqref{update}, we see that compared to letting $\rho_v(x) = 1$, our choice of $\rho_v(\cdot)$ is more likely to lead to a smaller $e_\mathrm{v}$, which can contribute to the reduction of $\zeta$, according to \eqref{23}. Moreover, to circumvent the singularity of $\rho_v(\cdot)$ at the origin, we let $\rho_v(x)= 1/(l^2(x,0)+ \rho)$, with a small positive constant $\rho$.
	
%	 Meanwhile, $\rho_v(x) = 1/l^2(x,0)$ is superior to $\rho_v(x) = 1$ in terms of approximation near the origin, as the former assigns higher weights for states around the origin. We have noticed in numerical experiments that systems are easily trapped in a fixed point that is not the origin if $\rho_v(x) = 1$. 
	
	\item Condition C2 requires the policy approximation error $\hat{J}_{\hat{\pi}_{\text {proj }}^{\mathrm{ex}}}(x)-\hat{J}_{\hat{\pi}^{\mathrm{im}}}(x)$ is constrained by the state cost $l(x, 0)$ scaled by a constant $e_\mathrm{p}$ and that the constant $e_\mathrm{p}$ is small enough. Based on this requirement, we take $\rho_\pi(x)=1 /(l(x, 0)+\rho)$ to make $e_\mathrm{p}$ small.
	
	\item C1 indicates that $\hat{J}_{k-1}(x) \geq 0,\;\forall x \in \Omega$ and $\hat{J}_{k-1}(0) =0$. To fulfill these, a modification of the PWA function approximator in \cite{dai2021lyapunov} can be adopted. In particular, for any PWA function approximator $\bar{J}(\cdot) : \mathbb{R}^{n_x} \to \mathbb{R}$, we can construct $\hat{J}_{i}(\cdot)$ via $\hat{J}_{i}(x) = \bar{J}(x)-\bar{J}(0) + |R_v x|_q$. Here, $R_v$ is a full column rank matrix that is included in the parameter $\theta$, and $q \in\{1, \infty\} $. Similarly, C2 implies that $\hat{\pi}^{\mathrm{ex}}(0)= 0$. To realize this, for any PWA function approximator $\bar{\pi}(\cdot) : \mathbb{R}^{n_x} \to \mathbb{R}^{n_u}$, we can let $\hat{\pi}^{\mathrm{ex}}(x)=\bar{\pi}(x)- \bar{\pi}(0)$.
	
	%	To interpret this, we find from \eqref{update} that the update of $\theta_{k}$ is for minimizing 
	%	\begin{equation}\label{update2}
		%		\mathbb{E}_{x\sim d}\left\{{{ \rho(x)( \Gamma_P \hat{J}_{k-1}(x) - \hat{J}_{k}(x) )}^2}\right\} 
		%	\end{equation}
	%	where $d$ is a distribution of states, related to the sampling strategy. If the states are sampled randomly, or from a uniform grid, 
	
\end{itemize}

\begin{remark}
Different from the existing stability results on ADP \cite{postoyan2019stability,moreno2022predictive,heydari2018stability,liu2017finite}, Theorem \ref{stable2} considers the effects of state constraints. It implies that constraint satisfaction can only be guaranteed in the sub-level set $\mathcal{B}(J^{\mathrm{soft}}_{k-1},\Omega) \cap \mathcal{B}(\hat{J}_{k-1},\Omega)$, which can be small, even if we have extremely small approximation errors. Besides, condition C1 allows us to analyze the sub-optimality of $\hat{\pi}^\mathrm{im}(\cdot)$ and $\hat{\pi}^\mathrm{ex}_\mathrm{proj}(\cdot)$, which has not been addressed in existing research \cite{postoyan2019stability,moreno2022predictive,heydari2018stability,liu2017finite}. 
\end{remark}
\subsection{Sub-optimality analysis}
Based on Theorem \ref{stable2}, we can compute upper bounds on $J_{\hat{\pi}^\mathrm{im}}(\cdot)$ and $J_{\hat{\pi}^\mathrm{ex}_\mathrm{proj}}(\cdot)$, which are the infinite-horizon costs of $\hat{\pi}^\mathrm{im}(\cdot)$ and $\hat{\pi}^\mathrm{ex}_\mathrm{proj}(\cdot)$, defined in \eqref{cost}.

\begin{corollary}\label{col}
	Consider Algorithm \ref{algo1} and the proposed policies $\hat{\pi}^\mathrm{im}(\cdot)$ and $\hat{\pi}^\mathrm{ex}_\mathrm{proj}(\cdot)$. Let $\Omega$ be a compact subset of $X$.
	
	(i) Let the assumptions in (i) of Theorem \ref{stable2} hold and $2\zeta \gamma <1$. Then, for any $x \in \mathcal{B}(J^{\mathrm{soft}}_{k-1},\Omega) \cap \mathcal{B}(\hat{J}_{k-1},\Omega)$, we have the inequality 
	\begin{equation}\label{performance1}
		J^{\mathrm{soft* }}(x) \leq J_{\hat{\pi}^\mathrm{im}}(x) \leq \frac{1-\zeta}{1-2\zeta \gamma} J^\mathrm{soft}_{k-1}(x).
	\end{equation}

(ii) Let the assumptions in (ii) of Theorem \ref{stable2} hold and $2\zeta \gamma + e_\mathrm{p}<1$. Then, for any $x \in \mathcal{B}(J^{\mathrm{soft}}_{k-1},\Omega) \cap \mathcal{B}(\hat{J}_{k-1},\Omega)$, we have the inequality 
\begin{equation}\label{performance2}
	J^{\mathrm{soft* }}(x) \leq J_{\hat{\pi}^\mathrm{ex}_\mathrm{proj}}(x) \leq \frac{1-\zeta}{1-2\zeta \gamma-e_\mathrm{p}} J^\mathrm{soft}_{k-1}(x).
\end{equation}
\end{corollary}

\textcolor{blue}{It is almost impossible to get an optimal control policy from \eqref{policy} due to the approximation error. However, \eqref{performance1} and \eqref{performance2} confirm the intuition that a smaller approximation error of the critic (and the actor) leads to tighter sub-optimality guarantees.} Namely, as $\zeta \to 0,\; e_\mathrm{p} \to 0$, and $ k \to \infty$, we have $J_{\hat{\pi}^\mathrm{im}}(x) \to J^{\mathrm{soft* }}(x)$ and $J_{\hat{\pi}^\mathrm{ex}_\mathrm{proj}}(x) \to J^{\mathrm{soft* }}(x)$ for any $x \in \mathcal{B}(J^{\mathrm{soft}}_{k-1},\Omega) \cap \mathcal{B}(\hat{J}_{k-1},\Omega)$. Moreover, if option (a) of Algorithm \ref{algo1} is used, $J^\mathrm{soft}_{k-1}(x)$ in \eqref{performance1} and \eqref{performance2} can be replaced by $J^{\mathrm{soft* }}(x)$ because $J^\mathrm{soft}_{k-1}(x) \leq J^\mathrm{soft*}(x),\;\forall x \in \mathcal{X}$.
\subsection{Stability and safety verification}

As mentioned in the previous subsection, conditions C1 and C2 in Theorem \ref{stable2} can only be evaluated statistically. Moreover, the conditions in Theorem \ref{stable2} are sufficient conditions for $\hat{J}_{k-1}(\cdot)$ and $J^\mathrm{soft}_{k-1}(\cdot)$ to be Lyapunov functions, and thus these conditions can be conservative. Besides, sometimes only practical stability can be ensured for nonlinear systems with neural controllers \cite{postoyan2019stability}. In this section, we propose an offline verification framework to simultaneously verify the practical stability and safety of the system controlled by the projected policy $\hat{\pi}^\mathrm{ex}_\mathrm{proj}(\cdot)$ in a deterministic manner, based on MILP. %The proposed framework directly evaluates the invariance of sub-level sets of $\hat{J}_{k-1}(\cdot)$, and also the set of trajectories of the closed-loop system in several time steps. By fixing some parameters, asymptotic stability can be further verified. 
A small adaption that can be used to verify the asymptotic stability for $\hat{\pi}^\mathrm{ex}_\mathrm{proj}(\cdot)$ and $\hat{\pi}^\mathrm{im}(\cdot)$ is provided at the end of this section.

The proposed verification procedure contains 3 steps. Different from \cite{chen2021learning,dai2021lyapunov,karg2020stability} that directly verify asymptotic stability, for practical stability we need to first verify the convergence of the closed-loop system to a neighborhood containing the origin, and then verify the invariance of the neighborhood. These will be formulated as two MILP problems. Finally, to enlarge the inner-estimated region of attraction, which is a sub-level set of $\hat{J}_{k-1}(\cdot)$, the third MILP problem will be formulated.

To verify the stability in any sub-level set $\mathcal{B}_{r_1} = \{x \in \mathcal{X}| \hat{J}_{k-1}(x)\leq r_1,\; r_1>0\}$ that is contained in $X$, we can directly extend the verifier in \cite{chen2021learning}, in which a mixed-integer quadratic program is solved to test a quadratic candidate Lyapunov function for PWA systems with input constraints. In particular, after implementing Algorithm \ref{algo1} and \eqref{policy}, we have the value function $\hat{J}_{k-1}(\cdot)$ and the explicit policy $\hat{\pi}^{\mathrm{ex}}\left(\cdot,\omega^*\right)$ at our disposal. Then, we formulate the following optimization problem:
\begin{align}\label{verify1}
	a^*_1 = \max _{x,u} &\; \hat{J}_{k-1}(f_\mathrm{PWA}(x,u))- \hat{J}_{k-1}(x) +c_1 l(x,0)  \nonumber\\
	\mathrm{s.t.} \;&  u = \hat{\pi}_{\mathrm{proj }}^{\mathrm{ex}}(x),\;r_2 \leq \hat{J}_{k-1}(x) \leq r_1,
\end{align}
where $c_1$ is a small positive parameter, and $r_2 \in (0,r_1)$. If $a^*_1 \leq 0$, we can conclude that the closed-loop system with $\hat{\pi}_{\mathrm{proj }}^{\mathrm{ex}}(\cdot)$ is safe in $\mathcal{B}_{r_1}$, and that any trajectories starting in $\mathcal{B}_{r_1}$ will enter $\mathcal{B}_{r_2} = \{x \in \mathcal{X}| \hat{J}_{k-1}(x)\leq r_2,\}$ in finite time. The formal result is included in Theorem \ref{theorem4}. If $a^*_1 >0$, we can reduce the values of $r_1$ and $c_1$. In this way, the objective function of problem \eqref{verify1} and the feasible region of $x$ become smaller, which will make $a^*_1$ smaller.

As all functions in \eqref{verify1} are PWA and thus MILP representable, problem \eqref{verify1} can be formulated as an MILP problem.

After the trajectories reach $\mathcal{B}_{r_2}$, we need to verify that they will always stay in $\mathcal{B}_{r_2}$, i.e., we need to prove the positive invariance of $\mathcal{B}_{r_2}$. This leads to the second MILP problem:
\begin{align}\label{verify2}
	a^*_2 = \max _{x,u} &\; \hat{J}_{k-1}(f_\mathrm{PWA}(x,u))- c_2 \hat{J}_{k-1}(x) -r_2+r_2 c_2  \nonumber\\
	\mathrm{s.t.} \;&  u = \hat{\pi}_{\mathrm{proj }}^{\mathrm{ex}}(x),\;0 \leq \hat{J}_{k-1}(x) \leq r_2,
\end{align}
where $c_2 \in [0,1]$. One can directly take $c_2 = 0$ to minimize $a^*_2$. Clearly, $a^*_2 \leq 0$ implies the positive invariance of $\mathcal{B}_{r_2}$, which will be proven in Theorem \ref{theorem4}. If $a^*_2 >0$, similarly we can make $r_2$ smaller.
%\begin{subequations}\label{verify2}
%\begin{align}
%	a^*_1 = \max _{x}\; & a \nonumber\\
%	\mathrm{s.t.} &  (x,u) = \mathrm{gr}_\mathrm{MILP}(\hat{\pi}^{\mathrm{ex}}) \label{verify2a} \\
%	&(u,u_\mathrm{proj}) = \mathrm{gr}_\mathrm{MILP}(\phi)\label{verify2b} \\ 
%	&([x^T\;u^T]^T,x^+) = \mathrm{gr}_\mathrm{MILP}(f_\mathrm{PWA}) \label{verify2c}\\ 
%	& (x,\varepsilon^{(j)}) = \mathrm{gr}_\mathrm{MILP}(\hat{J}_{k-1})\\
%	&  (x^+,\varepsilon^{(j)+}) = \mathrm{gr}_\mathrm{MILP}(\hat{J}_{k-1})\\
%	& (x, \varepsilon^{(l)}) = \mathrm{gr}_\mathrm{MILP}(l(\cdot,0)) \\
%	& ([c_1 \varepsilon^{(l)} - \varepsilon^{(j)}\;\; \varepsilon^{(j)+}- \varepsilon^{(j)}+ c_2 \varepsilon^{(l)}]^T,a) = \mathrm{gr}_\mathrm{MILP}(\max) \label{verify2g} \\ 
%	& \varepsilon^{(j)} \leq r
%\end{align}
%\end{subequations}
%where $u = \hat{\pi}_{\mathrm{proj }}^{\mathrm{ex}}(x)$ is converted into the constraints \eqref{verify2a} and \eqref{verify2b}, and the objective function in \eqref{verify1} is transformed into the constraints \eqref{verify2c}-\eqref{verify2g}.

However, the stable and safe region $\mathcal{B}_{r_1}$ derived from \eqref{verify1} and \eqref{verify2} is usually small, as $\mathcal{B}_{r_1}$ is a sub-level set of $\hat{J}_{k-1}(\cdot)$. In particular, if the weights in $Q$ on different states vary greatly, the resulting $\mathcal{B}_{r_1}$ will be rather narrow and much smaller than the real region of attraction. It could also happen that the evolution of the closed-loop system does not make $\hat{J}_{k-1}(\cdot)$ decrease at the beginning, but will drive the state into $\mathcal{B}_{r_1}$ in a finite number of time steps. Besides, in most cases we are interested in the performance of a policy in a polyhedron (or a UoP), rather than a sub-level set. 

Therefore, we further develop the third optimization problem that evaluates the range of trajectories of the closed-loop system in a finite number of time steps for all initial states in a polyhedron $X_\mathrm{in}$ or a UoP of interest. The problem is
\begin{align}\label{verify3}
	&\mathrm{Check}\; \mathrm{if}\; x_t \in X,\; t\!=\!1,...,N\!-\!1, \forall x_0 \in X_\mathrm{in} \nonumber\\
	& \mathrm{and}\; \mathrm{if}\; \hat{J}_{k-1}(x_N)\!\leq\! r_1, \forall x_0 \in X_\mathrm{in} \nonumber\\
	&\mathrm{s.t.} \;  u_t = \hat{\pi}_{\mathrm{proj }}^{\mathrm{ex}}(x_t),\;x_{t+1} = f_\mathrm{PWA}(x_t,u_t),\;t=0,...,N-1,
\end{align}
where $N$ is the number of time steps and $r_1$ is such that it makes $a^*_1 \leq 0$ in \eqref{verify1}. If \eqref{verify3} returns “Yes”, we can conclude that for any initial state in $X_\mathrm{in}$, the states of the closed-loop system will satisfy the constraints from $t=0$ to $t=N-1$, and the final state $x_N$ will reach the stable and safe region $\mathcal{B}_{r_1}$ computed in \eqref{verify1}. Similar to \eqref{verify1} and \eqref{verify2}, \eqref{verify3} can be exactly expressed in an MILP form. If $X_\mathrm{in}$ is a UoP, we also need some additional binary variables to formulate the initial condition $x_0 \in X_\mathrm{in}$ (see Appendix \ref{MILP formulation}).

The integration of \eqref{verify1}, \eqref{verify2}, and \eqref{verify3} constitutes the proposed verification framework, which computes the exact evolution of the closed-loop system. It does not need any sampling or statistical testing procedure. The effectiveness of the proposed verification framework is stated in the following theorem, of which the proof is provided in Appendix \ref{theorem4proof}.

\begin{theorem}\label{theorem4}
	Consider the policy $\hat{\pi}^\mathrm{ex}_\mathrm{proj}(\cdot)$ and the proposed verification framework consisting of \eqref{verify1}, \eqref{verify2}, and \eqref{verify3}. If $\hat{J}_{k-1}(0)=0$, $a^*_1 \leq 0$, $a^*_2 \leq 0$, and \eqref{verify3} returns “Yes”, then the closed-loop system with $\hat{\pi}^\mathrm{ex}_\mathrm{proj}(\cdot)$ is safe in $X_\mathrm{in}$, and any trajectory starting from $X_\mathrm{in}$ will approach $\mathcal{B}_{r_1}$ in at most $N + \lceil(r_1-r_2)\hat{\gamma}/(c_1 r_2)\rceil$ time steps and stay in $\mathcal{B}_{r_1}$ thereafter. Here $\hat{\gamma}$ is a positive constant independent of the initial condition.
\end{theorem}

With the results in Theorems \ref{stable2} and \ref{theorem4}, we now make practical suggestions on how to implement Algorithm \ref{algo1} and the proposed verification framework. 

After the explicit policy $\hat{\pi}^{\mathrm{ex}}\left(\cdot, \omega^*\right)$ is obtained, we first find $r_2$ that makes $a^*_2 \leq 0$. Then, one can use \eqref{verify2} to compute the safe and stable region and then use \eqref{verify3} to enlarge it. If $a^*_2 \leq 0$ but $a^*_1 >0$ whatever $r_1$ is chosen, one needs to apply \eqref{verify3} with $r_1$ replaced by $r_2$. The cost is that one may have to choose a large horizon $N$ since $r_2$ is small. The complexity of \eqref{verify3} grows exponentially w.r.t. $N$. If $a^*_1 > 0$ and $a^*_2 > 0$ no matter what $r_1$ and $r_2$ are chosen, one have to refine the learning process. To this end, Theorem \ref{stable2} implies that one can either (i) increase the number of iterations by tightening the stopping condition, i.e., making $e_{\text {tole }}$ smaller, or (ii) improve the approximation quality of function approximators and restart Algorithm \ref{algo1}.

\begin{remark}
The proposed verification framework generalizes the verification methods in \cite{schwan2022stability,chen2021learning,dai2021lyapunov,karg2020stability}, and is an extension of \cite{schwan2022stability,dai2021lyapunov,karg2020stability} from linear systems to PWA systems and from asymptotic stability to practical stability. Specifically, if we remove \eqref{verify2} and \eqref{verify3} and fix $r_2 = 0$, the proposed method verifies asymptotic stability, which is stronger than the properties in Theorem \ref{theorem4}, and it is then similar to the methods in \cite{chen2021learning,dai2021lyapunov}. For the comparison with \cite{karg2020stability}, we note that to guarantee state constraint satisfaction, \cite{karg2020stability} needs to search for a positively invariant set. In comparison, $X_\mathrm{in}$ in \eqref{verify3} is not necessarily positively invariant. Besides, \eqref{verify3} includes the case when the state constraint set $X$ and the searching region $X_\mathrm{in}$ are UoPs. Nevertheless, we should mention that similarly to \cite{schwan2022stability,karg2020stability}, one should refine the learning process as mentioned before and re-learn a policy if the verification scheme fails, while the approaches in \cite{chen2021learning,dai2021lyapunov} are more adaptive since they actively learn a Lyapunov function that can be different from $\hat{J}_{k-1}(\cdot)$.
\end{remark}

In addition, \eqref{verify1} can be adjusted to verify the asymptotic stability and safety of $\hat{\pi}^{\mathrm{im}}(\cdot)$. With $r_2 =0$ and $c_1l(x,0)$ replaced by $c_1l(x,u)$, $a^*_1 \leq 0$ tells that $\hat{J}_{k-1}(\cdot)$ is a control Lyapunov function, which further validates the asymptotic stability and safety of $\hat{\pi}^{\mathrm{im}}(\cdot)$ in $\mathcal{B}_{r_2}$. 

\subsection{Overall design procedure}
\begin{figure}
	\centering
	\includegraphics[width=230pt,clip]{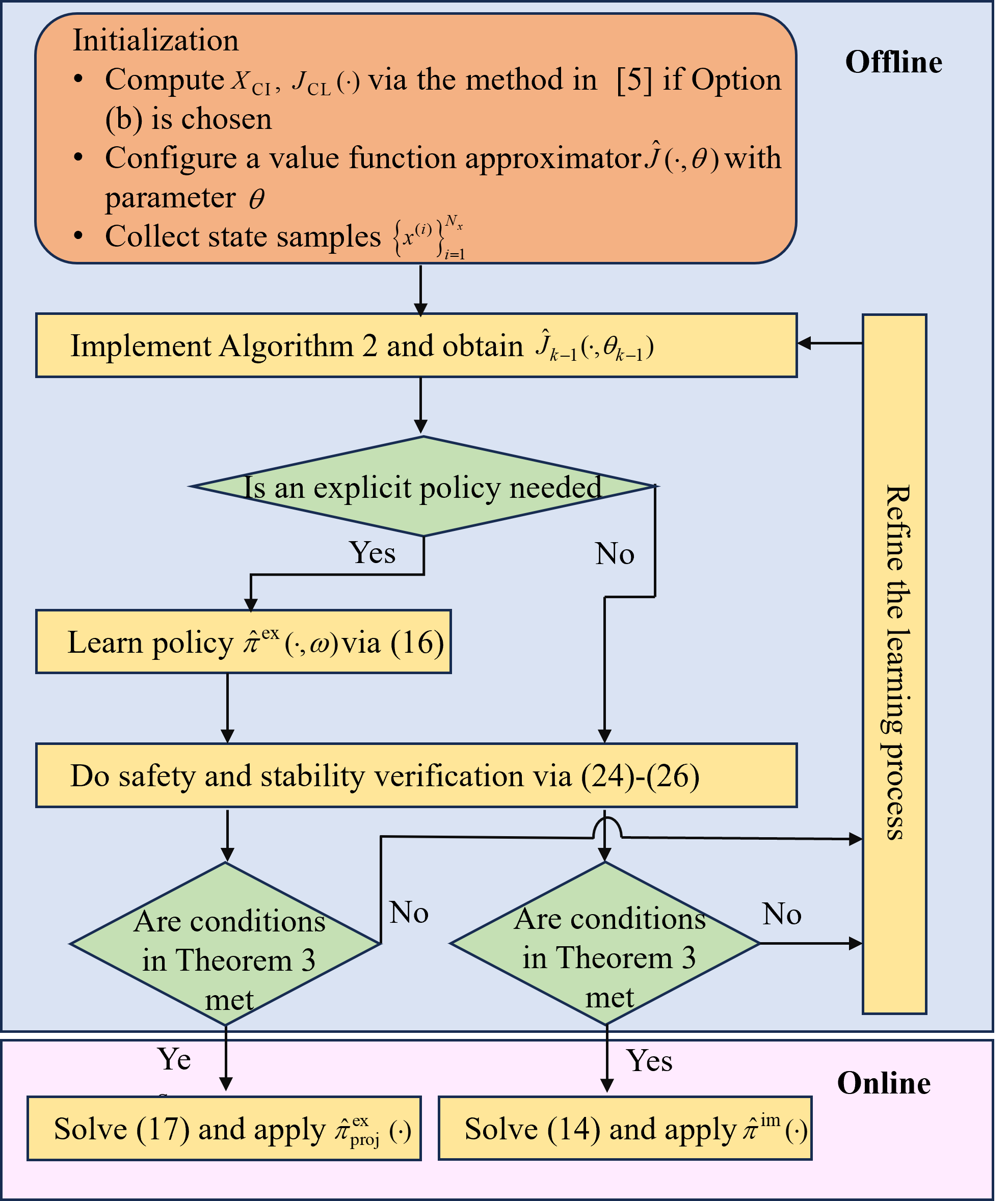}
	\caption{The schematic diagram of the design procedure}
\end{figure}
We provide a roadmap (see Fig. 1) outlining the procedures for implementing the proposed methods. The flowcharts within the blue box depict offline processes, while the flowcharts in the pink box represent online steps. After Algorithm \ref{algo1} is implemented, two options can be chosen. One can either learn the explicit policy via \eqref{policy} or bypass this step. Then, the proposed verification tool in Section V.C is used to assess the performance of the policy. If the policy falls short of expectations, it becomes necessary to refine the learning process and restart Algorithm \ref{algo1}. After verifying the safety and stability of the policy ${\hat \pi ^{{\text{ex}}}}( \cdot ,\omega )$ or ${\hat \pi ^{{\text{im}}}}( \cdot )$, in the online phrase we solve \eqref{project} or \eqref{suboptimal} to compute the control input $\hat{\pi}_{\text {proj }}^{\text {ex }}(x_t)$ or $\hat{\pi}^{\mathrm{im}}(x_t)$ when the state measure $x_t$ is received. 

The control framework involves four types of parameters: the hyper-parameters of the learning models for the value function and the policy, the penalty weight $p$, the parameter $e_{\text {tole }}$ of the tolerance function $\epsilon(\cdot)$ determining the stopping criterion of Algorithm \ref{algo1}, and the parameters $r_1$, $r_2$, and $N$ of the verification framework. Guidelines for setting hyper-parameters of the learning models can be found in machine learning literature, such as \cite{goodfellow2016deep}. The principle for tuning $r_1$, $r_2$, and $N$ has been provided after Theorem 3. To refine the learning process, one can adjust the first three kinds of parameters by the following steps: (i) improving the approximation quality of the learning models by tuning the hyper-parameters or amplifying samples; (ii) raising $p$; (iii) increasing the number of iterations by tightening the stopping condition ($e_{\text {tole }}$ smaller).

\section{Case study}
We validate the proposed methods in two examples. The simulations are conducted in MATLAB R2021a on an AMD Core R7-5800H CPU @3.20GHz machine. To guarantee a fair comparison of CPU times, all algorithms are implemented in MATLAB code. In particular, all MILP problems are solved by the MATLAB function “intlinprog” while all nonlinear problems are solved by the MATLAB function “fmincon” with the active-set algorithm.

\subsection{Example 1: inverted pendulum with elastic walls}
We consider a physical system, in which an active inverted pendulum is placed between elastic walls to maintain the vertical position (see Fig. 2(a)). The system parameters are chosen as $m = 1\;\mathrm{kg},\;l = 1\;\mathrm{m},\;d_1=0.12\;\mathrm{m},\;,d_2 = d_3 = 0.1\;\mathrm{m},\;k_1 = 300\; \mathrm{N/m},\;k_2 = 300\; \mathrm{N/m}$, and $k_3 = 500\; \mathrm{N/m}$. The state of the system is $x = [q\;\;\dot{q}]^T$. By linearizing the dynamics around the vertical configuration $q=\dot{q}=0$ and discretizing the system with a sampling time $0.05\;\mathrm{s}$, we obtain a discrete-time PWA model with 4 modes.
\begin{align*}
	A_{1}&=\left[\begin{array}{cc}1 & 0.05 \\ -29.5 & 1\end{array}\right], \quad B_{1}=\left[\begin{array}{c}0 \\ 0.05\end{array}\right], \quad f_{1}=\left[\begin{array}{l}0 \\ -3.3\end{array}\right]\\
	\mathcal{C}_1& = \{(x, u) \mid [1\;\;0]x \leq -0.12\}\\
	A_{2}&=\left[\begin{array}{cc}1 & 0.05 \\ -14.5 & 1\end{array}\right], \quad B_{2}=\left[\begin{array}{c}0 \\ 0.05\end{array}\right], \quad f_{2}=\left[\begin{array}{l}0 \\ -1.5\end{array}\right]\\
	\mathcal{C}_2& = \{(x, u)\mid -0.12 \leq  [1\;\;0]x \leq -0.1\}\\
	A_{3}&=\left[\begin{array}{cc}1 & 0.05 \\ 0.5 & 1\end{array}\right], \quad B_{3}=\left[\begin{array}{c}0 \\ 0.05\end{array}\right], \quad f_{3}=\left[\begin{array}{l}0 \\ 0\end{array}\right]\\
	\mathcal{C}_3& = \{(x, u)\mid -0.1 \leq   [1\;\;0]x \leq 0.1\}\\
	A_{4}&=\left[\begin{array}{cc}1 & 0.05 \\ -24.5& 1\end{array}\right], \quad B_{4}=\left[\begin{array}{c}0 \\ 0.05\end{array}\right], \quad f_{4}=\left[\begin{array}{l}0 \\ 2.5\end{array}\right]\\
	\mathcal{C}_4& = \{(x, u) \mid [1\;\;0]x \geq 0.1\}
\end{align*}

\begin{figure}\label{value_function}
	\centering
	\subfloat[An inverted pendulum with elastic walls.]{\includegraphics[width=120pt,clip]{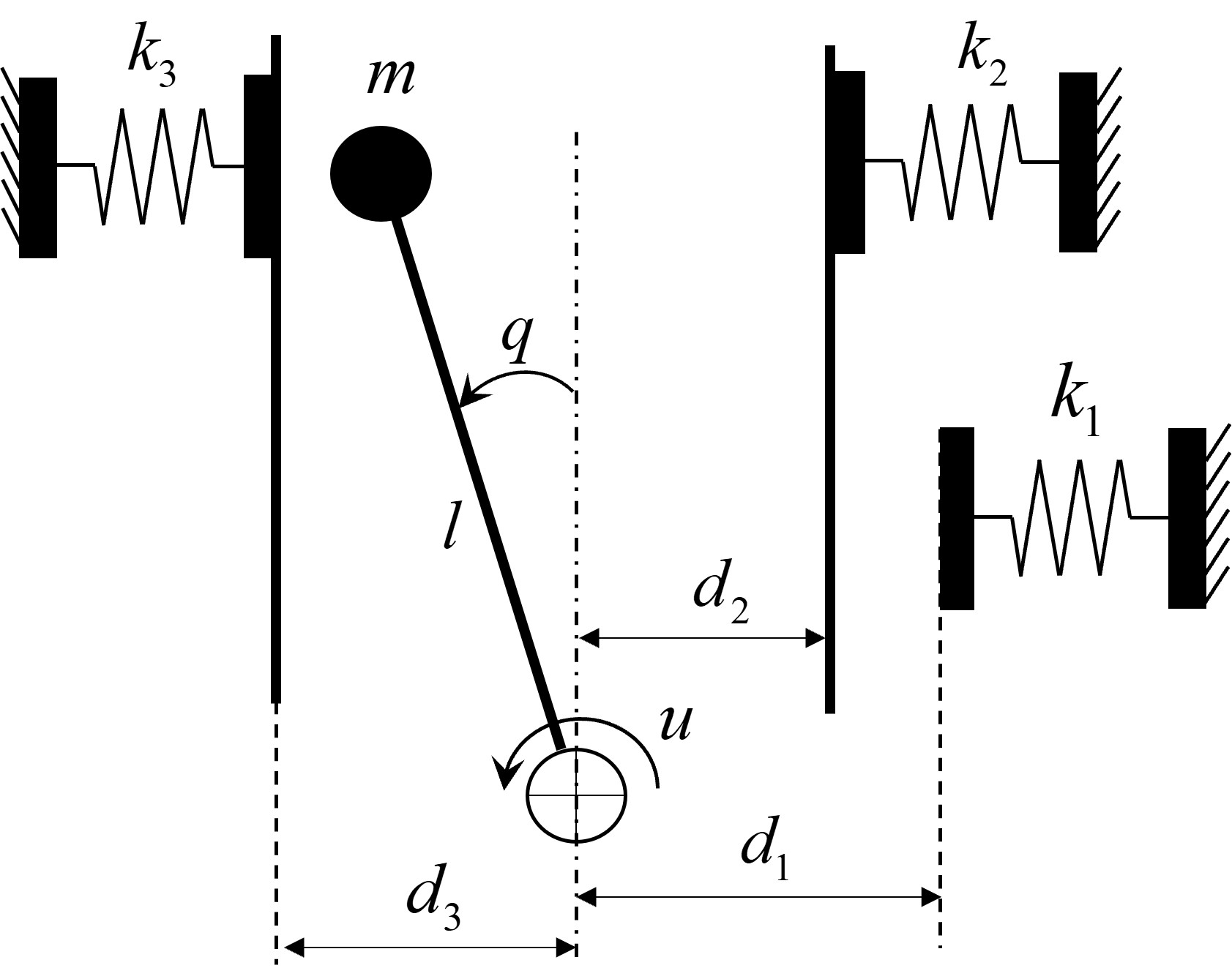}}
	\hfil
	\subfloat[The learning process of the ADP algorithm with different options.]{\includegraphics[width=120pt,clip]{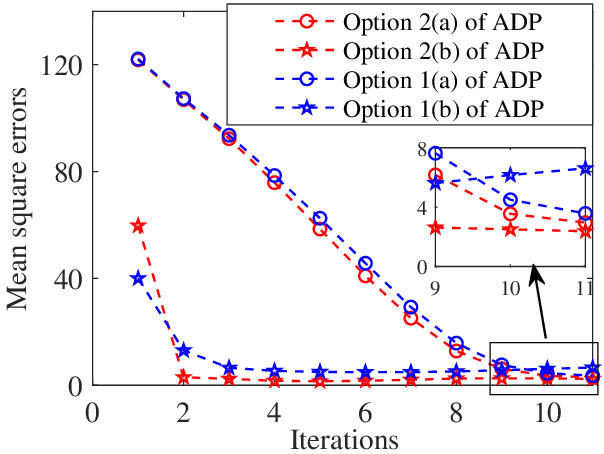}}
	\hfil
	\subfloat[Closed-loop trajectories of the inverted pendulum. The shaded region is $\bar{X}$.]{\includegraphics[width=120pt,clip]{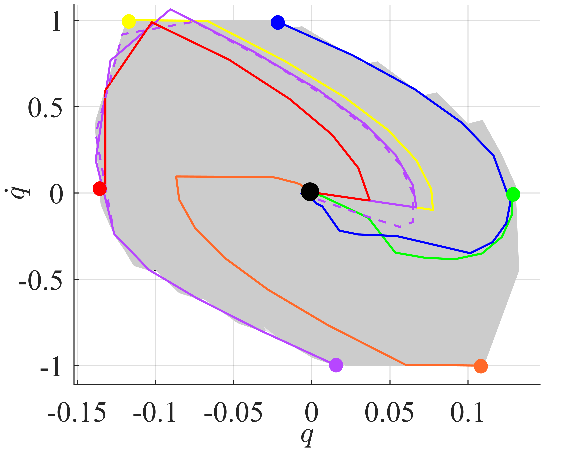}}
	\hfil
	\subfloat[Time-domain system responses.]{\includegraphics[width=120pt,clip]{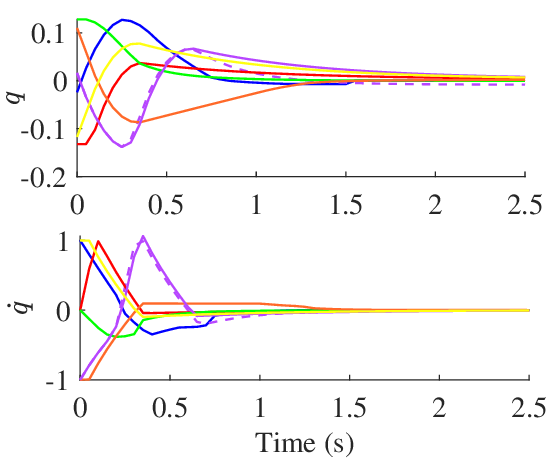}}
	\hfil
	\subfloat[Safe and stable regions verified by the verification framework.]{\includegraphics[width=120pt,clip]{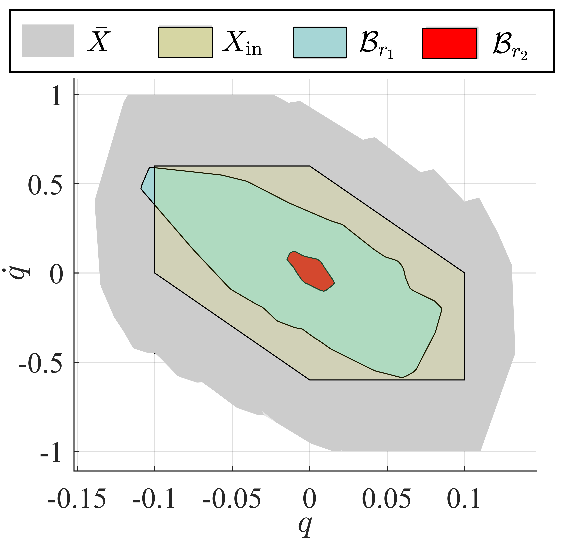}}
	\hfil
	\subfloat[Statistical analysis of different methods regarding CPU times. D-max means the difference of two max-affine functions.]{\includegraphics[width=120pt,clip]{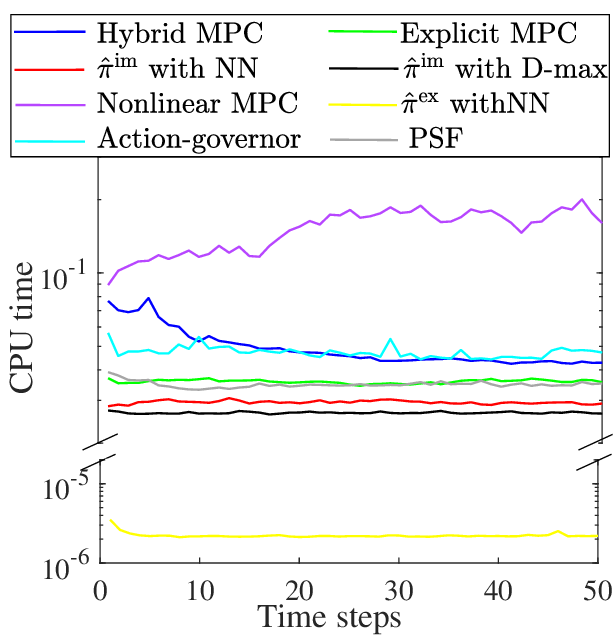}}
	\hfil
	\subfloat[Closed-loop trajectories of the inverted pendulum with UoP state constraints.]{\includegraphics[width=120pt,clip]{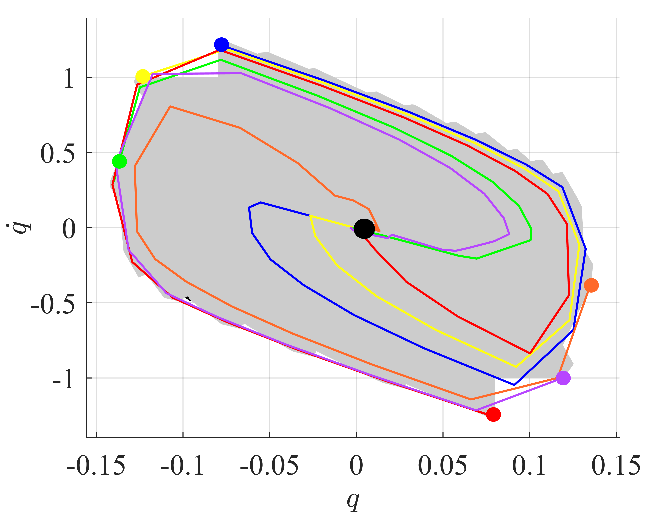}}
	\hfil
	\subfloat[Time-domain system responses under UoP state constraints.]{\includegraphics[width=120pt,clip]{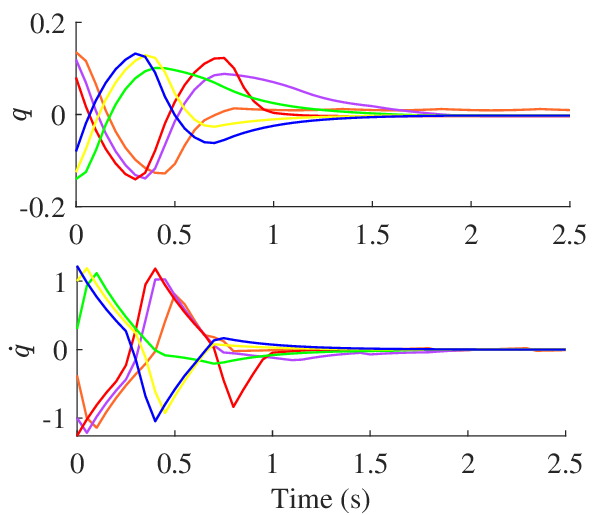}}
	\caption{Simulation results.}
\end{figure}
%\begin{figure}\label{nonconvex_fig}
%	\centering
%	\includegraphics[width=150pt, height=120pt,clip]{trajectory_nonconvex}
%	\caption{Closed-loop trajectories of the inverted pendulum with a UoP state constraint. The shaded region is $X_10$. The curves with different colors represent the state trajectories starting from different vertices.}
%\end{figure}

We can see that $f_1$, $f_3$, and $f_4$ are not zero. These affine terms are caused by the spring forces when the pendulum touches the elastic walls. Due to safety considerations and the limited capability of the torque motor, the inverted pendulum system is supposed to satisfy the constraints $[-0.15\;\;-1]^T\leq x \leq [0.15\;\;1]^T$ and $-4 \leq u \leq 4$. The constrained infinite-horizon control problem is constructed with $l(x,u)=|| \mathrm{diag}([20\;\;1])x||_\infty + ||u||_\infty$.  The overall control objective is to solve the infinite-horizon optimal control problem given in (1). The offline procedure for computing the proposed control policies $\hat{\pi}^{\mathrm{im}}$ and $\hat{\pi}^{\mathrm{ex}}$ includes solving several optimization problems. In particular, implementing Algorithm \ref{algo1} requires solving the MILP problem \eqref{target} and the nonlinear regression problem \eqref{update}. Computing $\hat{\pi}^{\mathrm{ex}}$ requires solving the policy optimization problem \eqref{policy}. Besides, the verification step contains solving three MILP problems \eqref{verify1}-\eqref{verify3}.

A 61 $\times$ 61 state data grid is constructed to cover the region $\{x|[-0.17\;\;-1.2]^T \leq x \leq [0.17\;\;1.2]^T\}$. These training points are exploited to train the actor and the critic. Both of them are ReLU NNs that have two hidden layers of widths 8. Starting from a zero value function and after 10 iterations, the closed-loop behavior of the system with $\hat{\pi}^\mathrm{ex}_\mathrm{proj}(\cdot)$ is illustrated in Figs. 2(c-d). Fig. 2(c) plots the trajectories of the closed-loop system (controlled by $\hat{\pi}^\mathrm{ex}_\mathrm{proj}(\cdot)$) starting from some vertices of $\bar{X}$ in the state space, while Fig. 2(d) displays the time-domain responses corresponding to the state-space trajectories of Fig. 2(c). Although states starting from these vertices are in general the most difficult to regulate to the origin, they converge rapidly under $\hat{\pi}^\mathrm{ex}_\mathrm{proj}(\cdot)$. Meanwhile, we should mention that the state constraints could be slightly violated: the trajectory depicted by the purple curves violates the constraints by about 5 percent. To avoid this, one can tighten the state constraints and restart the ADP algorithm. The dashed purple curve describes the trajectory starting from the purple vertex with 10 percent constraint tightening\footnote{We leverage a backtracking strategy to incrementally increase the constraint tightening factors until the safety performance is successfully verified by our proposed verification method.}. We can observe that constraint violation is avoided.

To carry out option (b) of Algorithm \ref{algo1}, we consider the linear subsystem $(A_3,\;B_3)$ and readily get a linear stabilizing feedback law $u_0 = [-40\;\; -10]x$, which is used to initialize the value function of option (b). The optimal value function $J^{*}(\cdot)$ is computed by the MPT3 toolbox \cite{MPT3} in 9.6 hours. The learning processes of different versions of Algorithm \ref{algo1} are compared in Fig. 2(b), in which the mean square errors between $\hat{J}_{k-1}\left(x^{(i)}, \theta_k\right)$ and  $J^{*}(x^{(i)})$ are depicted. One can see that the value function approximation in option (b) has a much faster convergence rate than that in option (a), and option 2 results in lower approximation errors than option 1. The accelerated convergence rate in option (b) is mainly attributed to its initial value function approximation $\hat{J}_0\left(\cdot, \theta_0\right)$ being considerably closer to $J^{\mathrm{soft} *}(\cdot)$ compared to option (a). The ultimately reduced approximation errors in option 2 likely stem from a more densely sampled state space.

The conditions for the stability and safety of the policy $\hat{\pi}_{\mathrm{proj}}^{\mathrm{ex}}(\cdot)$ are verified in Fig. 2(e). Fig. 2(e) depicts the safe and stable regions that are analyzed by the proposed verification framework. The input constraints are always satisfied in the simulation because the optimization problems \eqref{suboptimal} and \eqref{project} contain the input constraints as hard constraints. The blue region represents $\mathcal{B}_{r_1}$, which is computed by \eqref{verify1} with $r_1 = 18, \;c_1 = 0.1,\;r_2 =3$. We note that for some states in $\mathcal{B}_{r_2}$, which is colored red, the objective function in \eqref{verify1} becomes positive, so the verification method in \cite{dai2021lyapunov} fails. However, \eqref{verify2} outputs a negative $a^*_2$ with $c_2 = 0.1$, which means that any trajectory of the closed-loop system starting from $\mathcal{B}_{r_1}$ will reach in the neighborhood $\mathcal{B}_{r_2}$ containing the origin in finite number of time steps. Furthermore, the safe and stable region $\mathcal{B}_{r_1}$ is enlarged to $X_\mathrm{in}$ (the yellow region) by \eqref{verify3} with $N =3$. However, we observe that the verified safe and stable polytope $X_\mathrm{in}$ may be conservative compared to our trajectory simulation in Fig. 2(c). Suppose that $X_\mathrm{in}\triangleq \{x\in \mathbb{R}^{n_x}| E_{X_\mathrm{in}}x \leq g_{X_\mathrm{in}}\}$. The conservatism is primarily attributed to the naive choice of $E_{X_\mathrm{in}}$. Practical strategies to mitigate this conservatism involve using a UoP $X_\mathrm{in}$ and extending the horizon $N$. These two methods, on the other hand, will increase the complexity of the MILP problem \eqref{verify3}.

We compare the CPU time of running the ADP-based controllers, hybrid MPC, explicit MPC computed by the MPT3 toolbox, nonlinear MPC that uses the nonlinear model, action-governor RL in \cite{li2022robust}, and the predictive safety filter (PSF) in \cite{wabersich2021predictive}. With an MPC horizon 8, the implementation of hybrid MPC requires to solve an MILP problem that contains 94 continuous variables, 28 binary variables, 328 inequality constraints, and 21 equality constraints. In comparison, for $\hat{\pi}^\mathrm{im}(\cdot)$ with the value function approximator chosen as the ReLU NN (or the difference of two max-affine functions that has 10 and 3 terms in the first and second max blocks), the resulting MILP problem has 29(15) continuous variables, 20(7) binary variables, 94(73) inequality constraints, and 3(4) equality constraints. We only compare the time of computation that needs the knowledge of the current state, including constructing and solving optimization problems. For statistical analysis, we randomly select 100 initial states and run the system in 50 time steps\footnote{The number of time steps is set based on the fact that in our numerical experiments most scenarios converge to the origin after this number of time steps.}. The average computation time per time step of the different methods is shown in Fig. 2(f). The computation of $\hat{\pi}^\mathrm{ex}_\mathrm{proj}(\cdot)$ requires the least amount of CPU time, which is around $2.2\times 10^{-6}$ s per time step. Besides, $\hat{\pi}^\mathrm{im}(\cdot)$ performs better than hybrid MPC, explicit MPC, action-governor RL, and PSF, regarding online computation time, both when using the ReLU NN (about 0.028 s) or the difference of two max-affine functions (about 0.026 s) for value function approximation.

We further consider the case when the state constraints are a UoP, namely, $x \in X^{(1)}  \cup X^{(2)}$ with $X^{(1)} =\{x | [-0.15\;\;-1]^T\leq x \leq [0.15\;\;1]^T\}$ and $X^{(2)} =\{x | [-0.08\;\;-1.5]^T\leq x \leq [0.08\;\;1.5]^T\}$. This means a higher angular velocity is allowed when the pendulum is close to its vertical configuration. After implementing option 2(a) of Algorithm \ref{algo1} in 10 iterations, $\bar X$ is depicted in Fig. 2(g). The trajectories of the closed-loop system with $\hat{\pi}^\mathrm{ex}_\mathrm{proj}(\cdot)$ starting from some vertices of $\bar X$ are plotted in Figs. 2(g-h), from which one can find that the proposed scheme is still valid even when the state constraints are a UoP.

\subsection{Example 2: centralized adaptive cruise control}
We consider an adaptive cruise control problem in which we drive three vehicles (followers) to follow a leading vehicle (leader) in a highway environment. The control objective is to keep a desired distance
$d = 20$ m between each two adjacent vehicles. 

\begin{figure}\label{car}
	\centering
	\includegraphics[width=250pt,clip]{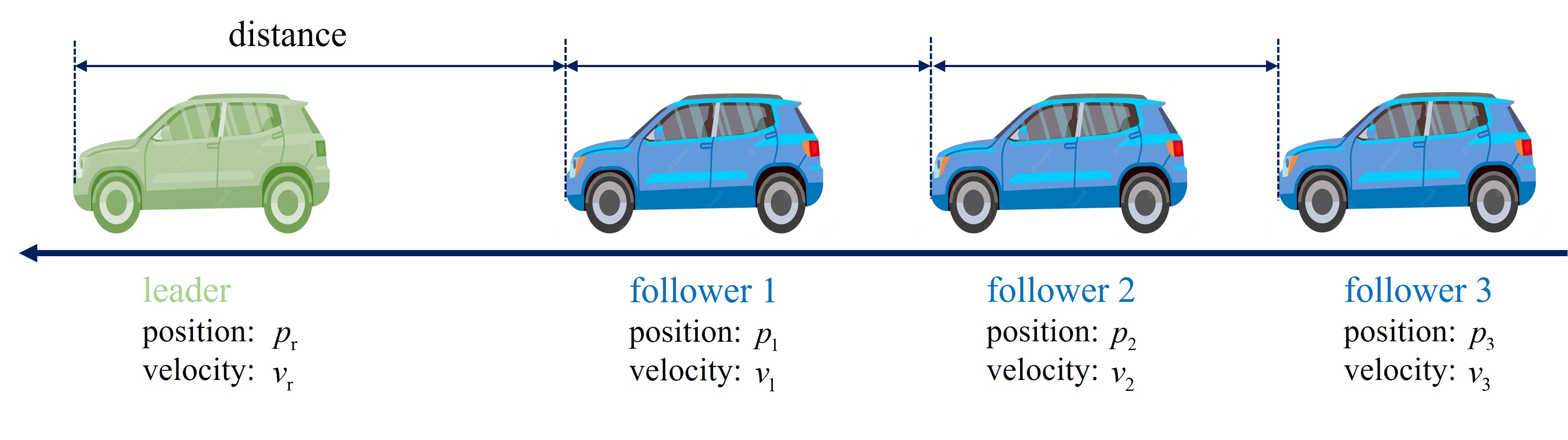}
	\caption{Adaptive cruise control set up.}
\end{figure}

\emph{Model:} The differential equation for the velocity $v_i(t),\;i=1,2,3$ of each following vehicle is $m \dot{v_i}(t)+c v^2_i(t)+\mu m g=b f_i(t)$, where $bf_i(t)$ is the input traction/brake force, and where $c =0.5 \; \mathrm{kg/m},\;m=800 \;\mathrm{kg},\;\mu = 0.01,\;b = 3700 \; \mathrm{N},\;g=9.8 \; \mathrm{m/s^2}$. The maximum allowable velocity for followers is $v_\mathrm{max} = 35 \;\mathrm{m/s}$.

Following \cite{corona2008adaptive}, a least-squares PWA approximation of the nonlinear damping force $V(v)=c v^2$ results in the continuous-time PWA dynamics with 2 modes for each vehicle
\begin{equation}\label{ex2}
	m \dot{v_i}(t)+c_j v_i(t)+a_j=b f_i(t),\; \mathrm{if}\; v_i(t) \in [\alpha_{j-1},\alpha_{j}), j=1,2
\end{equation}
where $\alpha_0 = 0,\;\alpha_1 = v_\mathrm{max}/2,\;\alpha_2 = v_\mathrm{max}$, $c_1 = 3c v_\mathrm{max}/8,\;c_2 = 13c v_\mathrm{max}/8$, $a_1 = \mu m g$, and $a_2 = -5c v_\mathrm{max}^2/8 + \mu m g$. The leader is assumed to have a constant velocity $v_\mathrm{r} = 20\;\mathrm{m/s}$. By defining a new state $x  = [p_1-p_\mathrm{r}-d\quad v_1-v_\mathrm{r}\quad p_2-p_1-d\quad v_2-v_\mathrm{r}\quad p_3-p_2-d \quad v_3-v_\mathrm{r} ]^T$, where $p_1,\;p_2,\;p_3$ and $p_\mathrm{r}$ are the positions of the three followers and the leader, respectively, the control goal is then making $x$ zero. The dynamics of $x$ thereby have $2^3$ different modes. Noticing that when $x$ is zero,  $v_i \in [\alpha_{1},\alpha_{2}), i=1,2,3$, so we introduce an input transformation $u = [f_1- \frac{v_\mathrm{r} c_2+a_2}{b} \quad f_2- \frac{v_\mathrm{r} c_2+a_2}{b} \quad f_3- \frac{v_\mathrm{r} c_2+a_2}{b}]^T$ to make the origin of $x$-$u$ space the equilibrium of the state equations for $x$. By discretizing the differential equations for $x$ with the sampling time 1s, we get a discrete-time PWA model with 6 states, 3 inputs, and 8 modes.

As for constraints, we consider limitations on the distance between each two adjacent vehicles, the velocity, and the traction/brake force input. In particular, we require that $ 10\;\mathrm{m} \leq |p_i-p_{i-1}| \leq 30\;\mathrm{m}$, $ 5 \;\mathrm{m/s} \leq v_i \leq  35 \;\mathrm{m/s}$, and $|f_i| \leq 1 \;\mathrm{N}$, where $i=1,2,3$ and $p_0 = p_\mathrm{r}$. All of these constraints can be readily converted into polyhedral constraints on $x$ and $u$.

The stage cost is chosen as $l(x, u)=\left\|\operatorname{diag}\left(\left[\begin{array}{llllll}
	1 & 0.5 & 1 &0.5 &1 &0.5
\end{array}\right]\right) x\right\|_{\infty}+\left\|\operatorname{diag}\left(\left[\begin{array}{lll}
0.1 & 0.1 & 0.1
\end{array}\right]\right) u\right\|_{\infty}$.

\emph{Setups:} We uniformly randomly sample 25000 state points from the state space $\{x \in \mathbb{R}^6\; |\; [-12\;\;-18\;\;-12\;\;-18\;\;-12\;\;-18]^T \leq x \leq [12\;\;18\;\;12\;\;18\;\;12\;\;18]^T\}$. We compare the performance of different controllers, including the proposed ADP controllers with different parameter settings, hybrid MPC with different prediction horizons, and the safety-filter-based RL in \cite{wabersich2021predictive}. The performance metrics include the averaged CPU time per time step, the cumulative stage cost over 15 time steps, and the percentage that the system trajectory satisfies all constraints. To test the performance of controllers, we generate 100 initial state scenarios that make the hybrid MPC with horizon 4 recursively feasible and stabilizing. Since controllable sets are difficult to compute in this example, we start from a zero value function and adopt option 2 of Algorithm \ref{algo1} to get $\hat{J}_{k-1}\left(\cdot, \theta_{k-1}\right)$. All PWA function approximators used in the proposed method and the safety-filter-based RL are ReLU NNs with 2 hidden layers.

\emph{Results:} Table 1 shows the results. In Table 1, $N$ represents the prediction horizon chosen in hybrid MPC, $M$ refers to the number of units in each hidden layer, and the safety rate is defined as the number of initial states leading to a safe closed-loop trajectory divided by the total number (100) of initial states. As expected, MPC provides the best performance regarding the total cost and safety rate, but it needs too much CPU time for computation. Compared with MPC, $\hat{\pi}^\mathrm{ex}_\mathrm{proj}$ and the method in  \cite{wabersich2021predictive} can reduce the online computation time, especially for $\hat{\pi}^\mathrm{ex}_\mathrm{proj}$. With the increase of the number of units, the performance of $\hat{\pi}^\mathrm{im}$ and $\hat{\pi}^\mathrm{ex}_\mathrm{proj}$ regarding the total cost becomes better and close to that of MPC. As both $\hat{\pi}^\mathrm{im}$ and $\hat{\pi}^\mathrm{ex}_\mathrm{proj}$ have acceptable performance regarding the total cost and safety rate, using $\hat{\pi}^\mathrm{ex}_\mathrm{proj}$ is preferable due to its significant reduction of CPU time. Besides, Table 1 also indicates that with the same number of units, the RL method in \cite{wabersich2021predictive} can be computed faster than $\hat{\pi}^\mathrm{im}$, but the performance of
$\hat{\pi}^\mathrm{im}$ and $\hat{\pi}^\mathrm{ex}_\mathrm{proj}$ is  superior to that of the RL method in \cite{wabersich2021predictive} in terms of both cumulative cost and constraint violations.

\begin{table}%
		\centering
		\caption{Performance of different controllers}%
		\begin{tabular}{p{70pt}p{50pt}p{40pt}p{40pt}}
			\toprule
			Methods&CPU time (mean/max) &Total cost & Safety rate \\
			\midrule
			MPC with $N=4$ & 0.1592 / 0.4593 & 21.7664 & 100\%\\
			MPC with $N=5$ & 0.2850 / 0.7703 & 21.6653 & 100\%\\
			MPC with $N=6$ & 0.4630 / 1.4619 & 21.5997 & 100\%\\
				\midrule
			$\hat{\pi}^\mathrm{im}$ with $M=15$ & 0.0605 / 0.1036 & 23.0097  & 98\% \\
			$\hat{\pi}^\mathrm{im}$ with $M=20$ & 0.0916 / 0.1621&  22.6613 & 100\%\\
			$\hat{\pi}^\mathrm{im}$ with $M=25$ & 0.1372 / 0.2991&  22.3042& 100\% \\
			$\hat{\pi}^\mathrm{im}$ with $M=30$ & 0.5487 / 1.3904& 22.0716 & 100\%\\
				\midrule
			$\hat{\pi}^\mathrm{ex}_\mathrm{proj}$ with $M=15$ & 0.0008 / 0.0028  & 23.4043 & 99\%\\
			$\hat{\pi}^\mathrm{ex}_\mathrm{proj}$ with $M=20$ & 0.0007 / 0.0027 & 22.9568 & 99\%\\
			$\hat{\pi}^\mathrm{ex}_\mathrm{proj}$ with $M=25$ & 0.0007 / 0.0027 & 22.7022 & 99\%\\
			$\hat{\pi}^\mathrm{ex}_\mathrm{proj}$ with $M=30$ & 0.0008 / 0.0030& 22.7562  & 99\%\\
				\midrule
			RL in \cite{wabersich2021predictive} with $M=30$,$N=5$ &  0.0809 / 0.3570 & 29.7561 & 85\%\\
			\bottomrule
		\end{tabular}
\end{table}

\section{CONCLUSIONS AND FUTURE WORK}

We have proposed an ADP control scheme to deal with infinite-horizon optimal control of PWA systems subject to linear and union-of-polyhedra (UoP) constraints, based on MILP. With carefully designed PWA penalty functions, the probably non-convex UoP constraints during the learning process are removed while the PWA properties of the value functions are maintained. We have formally analyzed the PWA properties and continuity of the value function, as well as the closed-loop stability and safety under the approximation errors. We have also designed an offline verification tool to make the proposed method reliable.  Simulation results show that the ADP-based policies are near-optimal, and require much less online computational effort than conventional hybrid MPC.

The limitations of the proposed ADP method are threefold. The performance of the policies depends heavily on the approximation accuracy of value functions. Secondly, our method does not scale well to high-dimensional problems due to the dramatic growth of sampling complexity. Additionally, our method is limited to the cases involving UoP state constraints and linear input constraints. Therefore, topics for future work include eliminating the reliance on value function approximation, exploring more efficient sampling strategies, and considering general convex multi-time-step constraints.

%%%%%%%%%%%%%%%%%%%%%%%%%%%%%%%%%%%%%%%%%%%%%%%%%%%%%%%%%%%%%%%%%%%%%%%%%%%%%%%%

\appendix
\section{Technical proofs}

\subsection{Proof of Theorem \ref{theorem1}}\label{theorem1proof}
The proof contains three parts.
\subsubsection{Convergence of the VI sequence in option 1}

\emph{Option 1(a):} For the nonlinear parametric problems in \eqref{algo3}, $U$ is closed, and the set $\left\{ {u \in U|\;J_k^{\mathrm{soft}} (x) = l(x,u) +P(x,X)+ J_{k - 1}^{\mathrm{soft}} \left( {{f_{{\text{PWA}}}}(x,u)} \right)} \right\}$ is nonempty for any $x$ in $\mathcal{X}$. Then, according to \cite[Theorem 4.2.1]{bank1983non}, it can be recursively proven that $J_{k}^{\mathrm{soft}}(\cdot),\;k=0,1,\dots$ is lower-semicontinuous on $\mathcal{X}$. Hence, all sub-level sets $\{x \in \mathcal{X} |J_{k}^{\mathrm{soft}}(x) \leq \lambda\}$ with $\lambda \in \mathbb{R}$ are closed. As $l_{p}(\cdot, \cdot)$ and $f_{\text {PWA }}(\cdot, \cdot)$ are continuous on $\mathcal{X} \times \mathcal{U}$, the set ${U_k}\left( {x,\lambda } \right) = \{ u \in U|\;{l }\left( {x,u} \right)+P(x,X) + J_{k - 1}^{\mathrm{soft}} \left( {{f_{{\text{PWA }}}}(x,u)} \right) \leq \lambda \}$ is closed and thus compact for all $x \in \mathcal{X}$, $\lambda  \in \mathbb{R}$, and $k \geq1$. As a result, the compactness assumption in \cite{bertsekas2015value} is satisfied and $\{J_{k}^{\mathrm{soft}}(\cdot)\}^\infty_{k=0}$ converges point-wise to $J^{\mathrm{soft* }}(\cdot)$, according to \cite[Proposition 2]{bertsekas2015value}. 

\emph{Option 1(b):} In this case, we define $J^{\mathrm{\infty}}_{0}(x) = \begin{cases}J_{\mathrm{CL}}(x), & x \in X_{\mathrm{CI}} \\ \infty & x \notin X_{\mathrm{CI}}\end{cases}$. The point-wise convergence of the VI $J^{\mathrm{\infty}}_{k}(x) =  \Gamma_{\mathrm{p,1}}J^{\mathrm{\infty}}_{k-1}(x)$ to $J^{\mathrm{soft} *}(x)$ is always guaranteed by \cite[Proposition 2]{bertsekas2015value} since $J^{\mathrm{\infty}}_{0}(x) \geq J^{\mathrm{soft} *}(x),\; \forall x \in \mathcal{X}$. %Meanwhile, it can be observed that $0\leq J^{\mathrm{soft}}_{0}(x)\leq J^{\mathrm{\infty}}_{0}(x),\; \forall x \in \mathcal{X}$. Then, as we have proven that the VI sequences starting from both zero and $J^{\mathrm{\infty}}_{0}(\cdot)$ converge point-wise to $J^{\mathrm{soft* }}(\cdot)$, 
By applying the monotonicity of the Bellman operator $\Gamma_{\mathrm{p,1}}$ \cite{bertsekas2019reinforcement}, we get the convergence of the VI sequence  $\{J_k^{\mathrm{soft }}(\cdot)\}^\infty_{k=0}$ to $J^{\mathrm{soft* }}(\cdot)$ in option 1(b).

\subsubsection{Continuity and PWA property of the value function in option 1}

By recursively iterating \eqref{algo3}, it is observed that $J_{k}^{\mathrm{soft}}(x)$ can be derived via the following batch approach:
\begin{align}\label{finite}
	J_{k}^{\mathrm{soft}}(x)=& \min _{u_{0}, \ldots, u_{k-1},\;{x_0}, \ldots ,{x_k}} \sum_{i=0}^{k-1} l_\mathrm{p}\left(x_{i}, u_{i}\right) +  J_0^{\mathrm{soft}}(x_k) \nonumber\\
	\text { s.t. } & x_{i+1}=f_{\text {PWA }}\left(x_{i}, u_{i}\right), x_{0}=x, \nonumber\\
	& u_{i} \in U, i=0, \ldots, k-1.
\end{align}

\emph{Option 1(a):} The proof of continuity follows from the proof of \cite[Corollary 17.2]{borrelli2017predictive}, because the objective function in \eqref{finite} is continuous and there is no state constraint. %In particular, problem \eqref{finite} is a multi-parametric program with only input constraints. The input constraints on $u_0,\;\dots,u_{k-1}$ are convex. By repeatedly substituting the state transition $x_{i+1}=f_{\mathrm{PWA}}\left(x_{i}, u_{i}\right),\;i=0,\dots,k-1$ into the objective function of \eqref{finite}, $\sum_{i=0}^{k-1} l_\mathrm{p}\left(x_{i}, u_{i}\right)$ is a continuous function of $x, u_0,\dots,u_{k-1}$. Therefore, the continuity of $J_{k}^{\mathrm{soft}}(x)$ follows from the continuity of $\sum_{i=0}^{k-1} l_\mathrm{p}\left(x_{i}, u_{i}\right)$, and \cite[Theorem 7]{hogan1973point}.

Now, we prove the PWA property of $J_{k}^{\mathrm{soft}}(\cdot)$. We denote by $\left\{\Xi _{j}\right\}_{j=1}^{r_0^{k}}$ the set of all possible switching sequences for $P(\cdot,\cdot)$, and by $\Xi_{j}^{(i)}$ the $i$-th element of the sequence $\Xi_j$. Namely, $\Xi_{j}^{(i)} =\vartheta  $ if $P(x_i,X) =  p  \max (0, (E^{(\vartheta)}_X)_{1\cdot}x  - (g^{(\vartheta)}_X)_{1}, \dots, (E^{(\vartheta)}_X)_{m^{(\vartheta)}_x\cdot}x - (g^{(\vartheta)}_X)_{m^{(\vartheta)}_x} )$. Here, without loss of generality we assume the first type of $P(\cdot,\cdot)$ in \eqref{penalty_define} is chosen. We notice that for a fixed switching sequence of $P(\cdot,\cdot)$, $P(\cdot,\cdot)$ can be converted into the linear inequality constraints $s^{(\vartheta)} \geq 0$ and $s^{(\vartheta)} \geq p(E^{(\vartheta)}_X)_{\ell\cdot}x  - p(g^{(\vartheta)}_X)_{\ell}$, $\ell = 1,...,m^{(\vartheta)}_x$ by introducing the slack variable $s^{(\vartheta)}$. Then, for a fixed switching sequence of the PWA dynamics and a fixed switching sequence of $P(\cdot,\cdot)$, problem \eqref{finite} reduces to a finite-horizon optimal control problem for a linear time-varying system with linear time-varying constraints, and thereby provides a PWA value function. As a result, $J_{k}^{\mathrm{soft}}(\cdot)$ is thus determined by comparing the value functions for all switching sequences and selecting the smallest one, so it is a PWA function in $\mathcal{X}$.

\emph{Option 1(b):} We firstly need to prove that the initial value function $J_{0}^{\mathrm{soft}}(\cdot)$ is continuous and PWA on $\mathcal{X}$. The continuity and PWA property of $J_{0}^{\mathrm{soft}}(\cdot)$ on $X_\mathrm{CI}$ is clear according to its definition in \eqref{soft2}. For problem \eqref{mplp}, there always exists a continuous optimizer $\bar z(\cdot)$ on $\mathcal{X}$, according to \cite[Theorem 6.5]{borrelli2017predictive}. Together with the continuity of $J_\mathrm{CL}(\cdot)$ and $P(\cdot,X_\mathrm{CI})$ in their domains, this ensures that $J_{0}^{\mathrm{soft}}(\cdot)$ is continuous on $\mathcal{X}\backslash  X_\mathrm{CI}$. The continuity of $J_{0}^{\mathrm{soft}}(\cdot)$ on the whole state space $\mathcal{X}$ thus follows from the fact that $\mathop {\lim }\limits_{x \to y,x \in \mathcal{X}/X_\mathrm{CI} } {J^\mathrm{soft}_0}\left( x \right) = {J_\mathrm{CL}}\left( y \right)$ holds for any $y \in \partial X_\mathrm{CI}$.
%To prove this, we notice that (i) $J_\mathrm{CL}(\cdot)$ is continuous on $X_\mathrm{CI}$, (ii) $\bar{z}(\cdot)$ is continuous on $\mathcal{X}$ and $\bar{z}(y)=y,\;\forall y \in X_0$, and (iii) $P(\cdot,X_\mathrm{CI})$ is continuous on $\mathcal{X}$ and $P(y,X_\mathrm{CI})=0,\;\forall y \in X_\mathrm{CI}$. A direct calculation shows that 
%\begin{align*}
%	\mathop {\lim }\tiny\limits_{\begin{array}{c}
%			x \to y\\x \!\in\! \mathcal{X}/X_\mathrm{CI}
%	\end{array}} \!\!{J^\mathrm{soft}_0}\left( x \right) &= \mathop {\lim }\tiny\limits_{\begin{array}{c}
%			x \to y\\x \!\in \!\mathcal{X}/X_\mathrm{CI}
%	\end{array}} \!\!J_\mathrm{CL}(\bar{z}(x))+P(x,X_\mathrm{CI})\nonumber\\
%&= J_\mathrm{CL}(y),\; \forall y \in \partial X_\mathrm{CI}
%\end{align*}

Besides, it is straightforward to prove that $J_{0}^{\mathrm{soft}}(\cdot)$ is also PWA outside $X_\mathrm{CI}$, since the optimizer $\bar z(\cdot)$ of the mp-LP, $J_\mathrm{CL}(\cdot)$, and $P(\cdot, X_\mathrm{CI})$ are all PWA \cite[Theorem 6.5]{borrelli2017predictive}. Due to the continuity and PWA property of $J_{0}^{\mathrm{soft}}(\cdot)$, the remainder of the proof of the continuity and PWA property of $J_{k}^{\mathrm{soft}}(\cdot)$ is similar to that in option 1(a). Thus, we have completed the proof of the statements of Theorem \ref{theorem1} in option 1.

\subsubsection{Equivalence of the VI sequences for options 1 and 2}
\textcolor{blue}{In option 2, by iterating $J_{i}^{\mathrm{soft}}(x) = \Gamma_{\mathrm{p,2}} J_{i-1}^{\mathrm{soft}}(x),\; \forall x \in \mathcal{X}$ from $i=1$ to $i=k$, we can get the expression of $J_k^{\mathrm{soft}}(\cdot)$ via the batch approach:}

\textcolor{blue}{\begin{align}\label{a1}
		&{J_k^{\mathrm{soft}}}\left( {{x}} \right) \nonumber\\
		=&\min _{u_0 \in U} \;l(x, u_0)+J_{k-1}^{\mathrm{soft}}\left(f_{\mathrm{PWA}}(x, u_0)\right)+ P_{k-1}\left(f_{\mathrm{PWA}}(x, u_0)\right)\nonumber\\
		=&\min _{u_0,u_1,x_0,x_1} l(x_0, u_0)+l(x_1, u_1)+ J_{k-2}^{\mathrm{soft}}\left(f_{\mathrm{PWA}}(x_1, u_1)\right)\nonumber\\
		&\quad\quad\quad\quad + P_{k-1}\left(x_1\right) + P_{k-2}\left(f_{\mathrm{PWA}}(x_1, u_1)\right)\nonumber\\
		&\quad  \mathrm{s.t.}\; x_0 =x,\;x_1 = f_\mathrm{PWA}(x_0,u_0),u_0,u_1\in U\nonumber\\
		=&\;...\nonumber\\
		=& \min _{\{u_{i}\}^k_{i=0},\;\{x_{i}\}^k_{i=0}} \sum_{i=0}^{k-1} \left(l\left(x_{i}, u_{i}\right)+ P_{k-1-i}\left(x_{i+1}\right)\right)  +  J_0^{\mathrm{soft}}(x_k) \nonumber\\
		&\text { s.t. }  x_{i+1}=f_{\text {PWA }}\left(x_{i}, u_{i}\right), x_{0}=x \nonumber\\
		& \quad \quad u_{i} \in U, i=0, \ldots, k-1.
	\end{align}
	Since $P_0(x) = 0, \forall x \in \mathcal{X}$ and $P_k(x) = P(x, X), \forall x \in \mathcal{X}$ and $\forall k>0$, by noticing that $l_\mathrm{p} (x,u) = l(x, u)+P(x, X)$ we have 
	\begin{align}\label{a2}
		{J_k^{\mathrm{soft}}}\left( {{x}} \right) =& \min _{\{u_{i}\}^k_{i=0},\;\{x_{i}\}^k_{i=0}} \sum_{i=0}^{k-1} l_\mathrm{p}\left(x_{i}, u_{i}\right) +  J_0^{\mathrm{soft}}(x_k) - P(x, X) \nonumber\\
		&\text { s.t. }  x_{i+1}=f_{\text {PWA }}\left(x_{i}, u_{i}\right), x_{0}=x \nonumber\\
		& \quad \quad u_{i} \in U, i=0, \ldots, k-1.	
	\end{align}
	By comparing the right-hand side of \eqref{a2} and (28), we can conclude that 
	\begin{align}\label{relation}
		{J_k^{\mathrm{soft}}}\left( {{x}} \right) = \mathrm{the\;optimal\;value\;of}\;(28) - P(x, X), \;\mathrm{for}\;k=0,1,...
	\end{align}
	From \eqref{relation}, we notice that the difference between the value functions in options 1 and 2 at the same iteration is $P(x, X)$, which is always continuous in $x$ and equals zero if $x \in X$. Combining \eqref{relation} with the first and second parts of the proof proves the statements of the theorem.}

\subsection{Proof of Lemma \ref{computex}}\label{lemma1proof}
The proof can be constructed by induction. Noticing that $X_{0}= X$ is a UoP, we assume that $X_{k-1}$ is a UoP in the form of $X_{k-1} = \bigcup_{i=1}^{r_{k-1}} X^{(i)}_{k-1}$. Then, the backward-reachable set $\mathrm{Pre}(X_{k-1})$ can be computed as $\mathrm{Pre}(X_{k-1}) =\bigcup_{i=1}^{r_{k-1}} \mathrm{Pre}(X^{(i)}_{k-1})$. This results in
\begin{align}\label{35}
	X_k& = \mathrm{Pre}(X_{k-1}) \cap X =  \bigcup_{j=1}^{r_{0}} \mathrm{Pre}(X_{k-1}) \cap X^{(j)} \nonumber\\
	& = \bigcup_{j=1}^{r_{0}} \left(\bigcup_{i=1}^{r_{k-1}} \mathrm{Pre}(X^{(i)}_{k-1})\right) \cap X^{(j)} \nonumber\\
	&= \bigcup_{j=1}^{r_0} \bigcup_{i=1}^{r_{k-1}} \mathrm{Pre}(X^{(i)}_{k-1})\cap X^{(j)}
\end{align}
In the last line of \eqref{35}, $\mathrm{Pre}(X^{(i)}_{k-1})\cap X^{(j)}$ is a UoP for each $i$ and $j$ because $X^{(i)}_{k-1}$ and $X^{(j)}$ are polyhedra. As a result, $X_k$ is a UoP. This proves the statement for $k$. Finally, by induction, we complete the proof of Lemma \ref{computex}.

\subsection{PWA functions and their mixed-integer formulations}\label{MILP formulation}
It has been shown that a ReLU NN can represent a PWA function \cite{montufar2014number}. A ReLU NN with $L$ hidden layers can be written by 
\begin{equation}\label{nn}
	f_\mathrm{NN}(x,\theta)=[f_{L+1} \circ f_\mathrm{ReLU} \circ f_{L} \circ \cdots \circ f_\mathrm{ReLU}  \circ f_{1}](x)
\end{equation}
where each hidden layer contains an affine map
$ f_{\ell}\left(\kappa_{\ell-1}\right)=W_{\ell} \kappa_{\ell-1}+b_{\ell}$, followed by a nonlinear map $f_\mathrm{ReLU}(f_\ell)= \max(0,f_\ell)$, $\ell=1,2,\dots,L$. Here, $M_\ell$ is the width of each layer, referring to the number of units in each layer, $\kappa_{\ell-1} \in \mathbb{R}^{M_{\ell-1}}$ is the output of the previous layer with $\kappa_{0}$ the input of the network, $W_\ell \in \mathbb{R}^{M_\ell\times M_{\ell-1}}$ and $b_\ell \in \mathbb{R}^{M_\ell}$ are the weights and biases, condensed in the parameter $\theta$.

Alternatively, \cite{kripfganz1987piecewise} has revealed that every continuous PWA scalar function can be written as the difference of two max-affine functions:
\begin{equation}\label{dmax}
	{f_{{\text{D}} - \max }}\left( {x,\theta } \right) = \max ({W^{(1)}}x + {b^{(1)}}) - \max ({W^{(2)}}x + {b^{(2)}})
\end{equation}
where $\max(a)$ selects the maximum element of the vector $a$, $W^{(1)} \in \mathbb{R}^{m_1 \times n_x}$, and where $W^{(2)} \in \mathbb{R}^{m_2 \times n_x}$, $b^{(1)} \in \mathbb{R}^{m_1}$ and $b^{(2)} \in \mathbb{R}^{m_2}$ constitute the parameter $\theta$. The regression method that uses the difference of two max-affine functions can thus be implemented by fixing the numbers $m_1,\;m_2$ of affine functions in the two max blocks and then updating $\theta$.

\emph{Mixed-integer formulation of PWA function approximators:} Consider the ReLU NN in \eqref{nn}, the outputs $\kappa_{\ell+1},\;\kappa_{\ell},\;\ell=0,1,\dots,L$ of any two adjacent layers satisfy $\kappa_{\ell+1}=f_\mathrm{ReLU}(W_{\ell} \kappa_{\ell}+b_{\ell})$. It has been shown in \cite{tjeng2017evaluating} that the ReLU function has the mixed-integer linear equivalent representation:
\begin{align*}
	&\mathrm{gr}_\mathrm{MILP}(f_\mathrm{ReLU})\nonumber\\
	=&\left\{(\kappa_{\ell},\kappa_{\ell+1}) \! \left| \! \begin{array}{l}
		\kappa_{\ell+1} \geq W_{\ell} \kappa_{\ell}+b_{\ell} \\
		\kappa_{\ell+1} \leq W_{\ell} \kappa_{\ell}+b_{\ell}\!-\!\operatorname{diag}\left(\underline{M}_{\ell}\right)\left({{\mathbf{1}}_{M \times 1}}\!-\!\delta_{\ell}\right) \\
		\kappa_{\ell+1} \geq 0 \\
		\kappa_{\ell+1} \leq \operatorname{diag}\left(\bar{M}_{\ell}\right) \delta_{\ell}
	\end{array}\right. \right\}
\end{align*}
where $\underline{M}_{\ell}$ and $\bar{M}_{\ell}$ are the lower and upper bounds of $W_{\ell} \kappa_{\ell}+b_{\ell}$, and $\delta_{\ell} \in \{0, 1\}^{M_\ell}$ is a binary variable. These bounds $\underline{M}_{\ell}$ and $\bar{M}_{\ell}$ can be computed by interval bound propagation or linear programming if the input of the network (the state of system \eqref{system}) is in a compact set \cite{weng2018towards}. Finally, by applying $\mathrm{gr}_\mathrm{MILP}(f_\mathrm{ReLU})$ to all the layers, one can derive the mixed-integer linear equivalent representation $\mathrm{gr}_\mathrm{MILP}(f_\mathrm{NN})$ of the NN \eqref{nn}.

For the approximator \eqref{dmax}, since \eqref{dmax} will be contained in the minimization problems \eqref{target} and \eqref{suboptimal}, the first max function can be simply converted into $m_1$ different linear inequalities. The second max function, according to \cite{tjeng2017evaluating}, leads to a mixed-integer formulation. In summary, we have
\begin{align}\label{mixed2}
	&\mathrm{gr}_\mathrm{MILP}(\max ({W^{(2)}}x + {b^{(2)}})) \nonumber\\
	=&\left\{(x,\varpi_2) \left|\begin{array}{l}
		\varpi_2 \geq W^{(2)}_{i,} x+b^{(2)}_i,\;\mathrm{for}\;i=1,\dots,m_2\\
		\varpi_2 \leq W^{(2)}_{i,} x+b^{(2)}_i + (1-\chi_i)(\bar{m}_{-i}-\underline{m}_i),\\
		 \quad \quad \quad \mathrm{for}\;i=1,\dots,m_2\\
		\sum\limits_{i = 1}^{m_2} {{\chi_i}}  = 1
	\end{array}\right. \right\}
\end{align}
where $\varpi_2$ and $\chi = [\chi_1\;\cdots\;\chi_{m_2}]^T$ are additionally introduced continuous and binary variables, respectively, $\underline{m}_i$ and $\bar{m}_{i}$ are the lower and upper bounds of $W^{(2)}_{i,} x+b^{(2)}_i$ in $X$, and $\bar{m}_{-i} = \max_{j  \ne i} \bar{m}_{j}$.

In addition, some other candidates for PWA approximation are, e.g., max-out NNs \cite{montufar2014number}, which inherently have equivalent mixed-integer formulations according to \eqref{mixed2}, and the PWA regression algorithm in \cite{bemporad2022piecewise}, in which a mixed-integer encoding of the algorithm is also provided.

\emph{Mixed-logical dynamical formulation of PWA systems:} In the traditional MPC framework for PWA systems \cite{lazar2006stabilizing}, the PWA model is usually converted into an equivalent mixed-logical dynamical form, so that the MPC problem is formulated as a solvable MILP problem. %More specifically, since $\mathcal{C}$ is compact, the original partition $\{\mathcal{C}_{i}\}^s_{i=1}$ of system \eqref{system} can always be transformed into a bounded partition after intersected with $\mathcal{C}$. Let $\{\tilde{\mathcal{C}}^{i}\}_{i=1}^{s} \triangleq \{\{\left[\begin{array}{ll}
	%	x^{T} & u^{T}
	%\end{array}\right]^{T} \in \mathbb{R}^{n_{x}+n_{u}}:{E_i}x + {F_i}u \leq {g_i}\}\}^s_{i=1} $ denote the new polyhedral partition of $\mathcal{C}$ derived by intersecting $\{\mathcal{C}_{i}\}^s_{i=1}$ with $\mathcal{C}$. 
	
	By specifying the partition $\left\{\mathcal{C}_{i}\right\}_{i=1}^{s}$ as $\{{\mathcal{C}}_{i}\}_{i=1}^{s} \triangleq \{\{\left[\begin{array}{ll}
		x^{T} & u^{T}
	\end{array}\right]^{T} \in \mathbb{R}^{n_{x}+n_{u}}:{\Psi^{(x)}_i}x + {\Psi^{(u)}_i}u \leq {\psi_i}\}\}^s_{i=1} $, according to \cite{borrelli2017predictive}, \eqref{system} can always be described as a mixed-logical dynamical (MLD) system. An equivalent representation \cite{bemporad1999control} is given by 
	\begin{align}\label{mld}
		&\mathrm{gr}_\mathrm{MILP}({f_{{\text{PWA}}}}) \nonumber\\
		=& \left\{\left(\left[\begin{array}{l}
			x \\
			u
		\end{array}\right], y\right) \left| \begin{array}{l}
			y = \sum\limits_{i = 1}^s {{z_i}} ,\;\sum\limits_{i = 1}^s {{\Delta _i}}  = 1 \hfill \\
			{\text{for}}\;i = 1,2,...,s: \hfill \\
			\left\{ \begin{gathered}
				{z_i} \leq M{\Delta _i} \hfill \\
				{z_i} \geq m{\Delta _i} \hfill \\
				{z_i} \leq {A_i}x + {B_i}u + {f_i} - m\left( {1 - {\Delta _i}} \right) \hfill \\
				{z_i} \geq {A_i}x + {B_i}u + {f_i} - M\left( {1 - {\Delta _i}} \right) \hfill \\
				{\Psi^{(x)}_i}x + {\Psi^{(u)}_i}u \leq {\psi_i} + M_i^*[1 - {\Delta _i}] \hfill \\ 
			\end{gathered}  \right. \hfill \\ 
		\end{array}  \right.\right\}
	\end{align}
	where $M$, $m$, and  $M_i^*$ are some bounds of affine functions of $x$ and $u$ over $X \times U$, and $\Delta _i \in \{0,1\}, i=1,2,\dots,s$ are binary variables. The detailed definitions of the bounds $M$, $m$, and  $M_i^*$ can be found in \cite{bemporad1999control}.
	
	The last inequality in \eqref{mld} is used to formulate the logic $[x^T\;\;u^T]^T \in \cup^s_{i=1} \mathcal{C}_i$. This can be generalized to convert any UoP constraint into a mixed-integer constraint, such as $x_0 \in X_{\mathrm{in }}$ in \eqref{verify3}.
	
\emph{Mixed-integer formulation of penalty functions in \eqref{penalty_define}:} As \eqref{target} and \eqref{suboptimal} are all minimization problems, the maximum operation in $P(\cdot,\cdot)$ can be eliminated by introducing some linear inequalities and continuous variables. On the other hand, the minumum operation in $P(\cdot,\cdot)$ results in a mixed-integer formulation similar to \eqref{mixed2}.

\subsection{Proof of Lemma \ref{lemma2}}\label{lemma2proof}

%	The proof is similar to the stability analysis of MPC \cite[Theorem 12.2]{borrelli2017predictive}, by showing that $J_{k}^{+}$ is a Lyapunov function.

Firstly, the result in (i) is straightforward, based on the fact that both $J_k^{\mathrm{soft}}(\cdot)$ and $l(\cdot,0)$ are continuous PWA functions on $\bar{X}$ and equal to zero iff $x =0$. 

For (ii), we will prove by contradiction the existence of a finite $\bar{k}$ such that  $J_{\bar{k}}^{\mathrm{soft}}(x)-J_{\bar{k}-1}^{\mathrm{soft}}(x) \leq \beta l(x,0),\;\forall x \in \bar{X} $ holds in option 1. Suppose that it does not hold, then for any $k>0$, there exists a $\bar{x} \in \bar{X} $ such that $J_k^{\mathrm{soft}}(\bar x)-J_{k-1}^{\mathrm{soft}}(\bar x) \geq  l(\bar x,0)$. According to the contraction property of the Bellman operator \cite{bertsekas2019reinforcement}, we have $J_i^{\mathrm{soft}}(\bar x)-J_{i-1}^{\mathrm{soft}}(\bar x) \geq  l(\bar x,0),\; \forall i \in \{1,...,k\}$. Summing over $i$ yields $J_k^{\mathrm{soft}}(\bar x) \geq  kl(\bar x,0)$. Letting $k \to \infty$ makes $J_\infty^{\mathrm{soft}}(\bar x)$ contradict (i) of Lemma \ref{lemma2}. Moreover, again using the contraction property of $\Gamma_{\mathrm{p,1}}$ leads to the conclusion in (ii). Finally, \eqref{relation} tells that the difference of $J_{\bar{k}}^{\mathrm{soft}}(x)$ in options 1 and 2 is $P(x,X)$, which is independent of $k$. Therefore, \eqref{stable} also holds for option 2 of Algorithm 1.
\subsection{Proofs of Theorem \ref{stable2} and Corollary \ref{col}}
For each $k \geq \bar{k}$, we define the policy $\pi_k(\cdot)$ as (one of) the optimizer(s) of $\Gamma_{\mathrm{p}, \alpha}J_{k-1}^{\mathrm{soft}}(\cdot),\;\alpha=1\;\mathrm{or}\;2$. For every $x \in \mathcal{B}( J^{\mathrm{soft}}_{k-1},\Omega) \subseteq  X$, according to \eqref{stable2},
\begin{equation}\label{32}
	J_{k}^{\mathrm{soft}}(x) = \Gamma_{\mathrm{p}, \alpha} J_{k-1}^{\mathrm{soft}}(x) \geq l(x, \pi_k(x)) + J_{k-1}^{\mathrm{soft}}(f_\mathrm{PWA}(x,\pi_k(x)))
\end{equation}
holds in both options. Together with \eqref{stable}, \eqref{32} yields
\begin{align}\label{33}
	J_{k-1}^{\mathrm{soft}}(f_\mathrm{PWA}(x,\pi_k(x))) - J_{k-1}^{\mathrm{soft}}(x) &\leq -(1-\beta) l(x,\pi_k(x)) \nonumber\\
	&\leq -(1-\beta) l(x,0),
\end{align}
which means that for every $x \in \mathcal{B}( J^{\mathrm{soft}}_{k-1},\Omega)$, the policy $\pi_k(\cdot)$ will make $f_\mathrm{PWA}(x,\pi_k(x)) \in \mathcal{B}( J^{\mathrm{soft}}_{k-1},\Omega)$.

Now, we show that $J^\mathrm{soft}_{k-1}(\cdot)$ and $\hat{J}_{k-1}(\cdot)$ are Lyapunov functions for the system with $\hat{\pi}^\mathrm{im}(\cdot)$. In particular, for any $x \in \mathcal{B}( J^{\mathrm{soft}}_{k-1},\Omega) \cap \mathcal{B}( \hat{J}_{k-1},\Omega) $, if C1 and \eqref{zeta11} hold, we have
	\begin{align}\label{fangsuo}
				&l(x, \hat{\pi}^\mathrm{im}(x)) + \hat{J}_{k-1}(f_\mathrm{PWA}(x, \hat{\pi}^\mathrm{im}(x))) \nonumber\\
				\leq &l(x, \pi_k(x)) + \hat{J}_{k-1}(f_\mathrm{PWA}(x,\pi_k(x))) \nonumber\\
				\leq &l(x, \pi_k(x)) + (1+\zeta)J^\mathrm{soft}_{k-1}(f_\mathrm{PWA}(x,\pi_k(x))) \nonumber\\
				\leq& (1+\zeta)J^\mathrm{soft}_{k-1}(x) + \left(1- (1+\zeta)(1-\beta)\right) l(x, \pi_k(x)) \nonumber\\
				\leq& \hat{J}_{k-1} (x) +2\zeta J^\mathrm{soft}_{k-1}(x) + \left(1- (1+\zeta)(1-\beta)\right) l(x, \pi_k(x)).
			\end{align}
In \eqref{fangsuo}, the first inequality is true since $\hat{\pi}^\mathrm{im}(\cdot)$ is an optimizer of problem \eqref{suboptimal}; the second and the last inequalities hold because $x$ and $f_\mathrm{PWA}(x,\pi_k(x))$ all in $\Omega$, in which C1 holds; and the third inequality is correct owning to \eqref{33}. 
Since $1-(1+\zeta)(1-\beta) <0$, considering Lemma \ref{lemma2}, \eqref{fangsuo} results in
\begin{align}\label{decrease}
	&\hat{J}_{k-1}(x^+) - \hat{J}_{k-1} (x) \leq \left(2\zeta \gamma- (1+\zeta)(1-\beta)\right)l(x,0),\nonumber\\
	&\mathrm{with}\;x^+ = f_\mathrm{PWA}(x, \hat{\pi}^\mathrm{im}(x)).
\end{align}
The right-hand side of \eqref{decrease} is strictly negative except for $x=0$ according to \eqref{zeta11}. Therefore, we can conclude that $\forall x \in \mathcal{B}( J^{\mathrm{soft}}_{k-1},\Omega) \cap \mathcal{B}( \hat{J}_{k-1},\Omega) $, we have $f_\mathrm{PWA}(x, \hat{\pi}^\mathrm{im}(x)) \in \mathcal{B}( \hat{J}_{k-1},\Omega)$. Meanwhile, condition C1 also implies that $\hat{J}_{k-1} (x) > 0,\;\forall x \in (\mathcal{B}( J^{\mathrm{soft}}_{k-1},\Omega) \cap \mathcal{B}( \hat{J}_{k-1},\Omega)) \backslash \{0\} $ and $\hat{J}_{k-1} (0)=0$. This shows that $J^\mathrm{soft}_{k-1}(\cdot) $  is a Lyapunov function for the closed-loop system $x_{t+1} = f_\mathrm{PWA}(x_t,\hat{\pi}^\mathrm{im}(x_t))$.

Similarly to \eqref{fangsuo} and \eqref{decrease}, we can get the following inequality for $J^\mathrm{soft}_{k-1}(\cdot)$: 
\begin{align}\label{fangsuo2}
	&l(x, \hat{\pi}^\mathrm{im}(x)) + J^\mathrm{soft}_{k-1}(f_\mathrm{PWA}(x,\hat{\pi}^\mathrm{im}(x))) \nonumber\\
	\!\leq\! & \frac{(1\!+\!\zeta) J^\mathrm{soft}_{k-1}(x) \!+\! (1\!-\!(1\!+\!\zeta)(1\!-\!\beta))l(x,\!\pi_k(x))\!-\!\zeta l(x,\!\hat{\pi}^\mathrm{im}(x))}{1-\zeta} .
\end{align}
With \eqref{zeta11} and Lemma \ref{lemma2}, \eqref{fangsuo2} readily leads to 
\begin{align}\label{decreasing}
	J^\mathrm{soft}_{k-1}(x^+)-J^\mathrm{soft}_{k-1} (x)  \leq \frac{2\zeta \gamma - (1+\zeta)(1-\beta)}{1-\zeta}l(x,0) ,
\end{align}
which means that $J^\mathrm{soft}_{k-1}(\cdot) $ is strictly decreasing from any $x \in (\mathcal{B}( J^{\mathrm{soft}}_{k-1},\Omega) \cap \mathcal{B}( \hat{J}_{k-1},\Omega)) \backslash \{0\} $ to the next state. By combining \eqref{decrease} and \eqref{decreasing}, we note that $\mathcal{B}( J^{\mathrm{soft}}_{k-1},\Omega) \cap \mathcal{B}( \hat{J}_{k-1},\Omega)$ is a positively invariant set for the closed-loop system. This together with the Lyapunov functions $\hat{J}_{k-1}(\cdot) $ and  $J^\mathrm{soft}_{k-1}(\cdot) $ leads to the asymptotic stability and safety of the closed-loop system.

Next, we analyze the behavior of the closed-loop system with $\hat{\pi}^\mathrm{ex}_\mathrm{proj}(\cdot)$. With C1 and C2, $\hat{J}_{k-1} (x^+)- \hat{J}_{k-1} (x)$ with $x^+ = f_\mathrm{PWA}(x,\hat{\pi}^\mathrm{ex}_\mathrm{proj}(x))$ is then upper bounded by the right-hand side of \eqref{decrease} plus $e_\mathrm{p}l(x,0)$, and $J^\mathrm{soft}_{k-1} (x^+)-J^\mathrm{soft}_{k-1} (x)$ is also upper bounded by the right-hand side of \eqref{decreasing} plus $e_\mathrm{p}l(x,0)$. Together with \eqref{zeta12} this results in the asymptotic stability and safety of the closed-loop system with policy $\hat{\pi}^\mathrm{ex}_\mathrm{proj}(\cdot)$.

Finally, to prove Corollary \ref{col}, we note that the first inequality in \eqref{performance1} and the first inequality in \eqref{performance2} directly follow from the optimality of $J^\mathrm{soft* }(\cdot)$. We derive from \eqref{fangsuo2} that 
\begin{equation}\label{39}
	\frac{1-2\zeta \gamma}{1-\zeta} l(x,\hat{\pi}^\mathrm{im}(x)) \leq  J^\mathrm{soft}_{k-1} (x)-J^\mathrm{soft}_{k-1}(f_\mathrm{PWA}(x, \hat{\pi}^\mathrm{im}(x))) 
\end{equation}
holds for any $x \in (\mathcal{B}( J^{\mathrm{soft}}_{k-1},\Omega) \cap \mathcal{B}( \hat{J}_{k-1},\Omega)) \backslash \{0\} $. Consider the trajectory $x_0, x_1, ...$ that is generated by applying $ \hat{\pi}^\mathrm{im}(x_t)$ to system \eqref{system} at each time step $t,\;t=0,1,...$ Letting $x=x_t$ and summing up both sides of \eqref{39} from $t=0$ to $t = \infty$, we get $ J_{\hat{\pi}^\mathrm{im}}(x_0) \leq \frac{1-\zeta}{1-2\zeta \gamma} (J^\mathrm{soft}_{k-1} (x_0)-J^\mathrm{soft}_{k-1} (x_\infty))$. The asymptotic stability in (i) of Theorem \ref{stable2} indicates that $J^\mathrm{soft}_{k-1} (x_\infty) = 0$, so (i) of Corollary \ref{col} is proved. Similarly, we can upper bound the stage cost when applying $\hat{\pi}^\mathrm{ex}_\mathrm{proj}(\cdot)$ by
\begin{align*}
	&\frac{1-2\zeta \gamma-e_\mathrm{p}}{1-\zeta} l(x,\hat{\pi}^\mathrm{ex}_\mathrm{proj}(x)) \nonumber\\
	\leq&  J^\mathrm{soft}_{k-1} (x)-J^\mathrm{soft}_{k-1}(f_\mathrm{PWA}(x, \hat{\pi}^\mathrm{ex}_\mathrm{proj}(x))) .
\end{align*}
(ii) of Corollary \ref{col} will be obtained by summing up the above inequality along the trajectory controlled by $\hat{\pi}^\mathrm{ex}_\mathrm{proj}(\cdot)$.
\subsection{Proof of Theorem \ref{theorem4}}\label{theorem4proof}
If \eqref{verify3} returns “Yes”, the trajectories of the closed-loop system with $\hat{\pi}^\mathrm{ex}_\mathrm{proj}(\cdot)$ with initial condition $x_0 \in X_\mathrm{in}$ will be contained in $\mathcal{B}_{r_1}$ after $N$ time steps. 

Then, for any initial state $x_0 \in \mathcal{B}_{r_1}$, suppose that the $t_\mathrm{f}-1$-step trajectory $x_0,\;x_1,...,\;x_{t_\mathrm{f}-1}$ of the closed-loop system with $\hat{\pi}^\mathrm{ex}_\mathrm{proj}(\cdot)$ is not contained in $\mathcal{B}_{r_2}$. Since $a^*_1 \leq 0$, we have 
\begin{equation}\label{41}
	\hat{J}_{k-1}(x_{t+1})-\hat{J}_{k-1}(x_t) \leq -c_1 l(x_t,0),\;t=0,...,t_\mathrm{f}-1
\end{equation}
Summing \eqref{41} over time yields
\begin{equation}\label{42}
	\hat{J}_{k-1}(x_{t_\mathrm{f}})\leq \hat{J}_{k-1}(x_0)  -c_1 \sum_{t=0}^{t_\mathrm{f}-1} l(x_t,0)
\end{equation}

Meanwhile, since both $\hat{J}_{k-1}(\cdot)$ and $l(x,0)$ are continuous PWA functions on their domains, similarly to (i) of Lemma \ref{lemma2}, there exists a positive and finite constant $\hat{\gamma}$ such that $\hat{J}_{k-1}(x) \leq \hat{\gamma} l(x,0)$ for all $x \in  X$. %To prove this, we see that $\hat{J}_{k-1}(x)/l(x,0)$ can only approach infinity when $x \to 0$. When $x$ is in a small neighborhood containing the origin, $\hat{J}_{k-1}(x)$ is linear in $x$ since $\hat{J}_{k-1}(0) =0$. Then we have $\frac{\hat{J}_{k-1}(x)}{l(x,0)} =\frac{W(x)x}{\|Qx\|_q} \leq \frac{\|W(x)Q^{-1}\|_q \|Q^{-1}x\|_q }{\|Qx\|_q}\leq  \|W(x)Q^{-1}\|_q$, where $W(x) \in \mathbb{R}^{1\times n_x}$. This confirms the existence of such $\hat{\gamma}$. 
As a result, \eqref{42} implies that $\hat{J}_{k-1}(x_{t_\mathrm{f}})\leq r_1 - \frac{c_1 r_2 t_\mathrm{f}}{\hat{\gamma}}$. Specifying $t_\mathrm{f}= \lceil(r_1-r_2)\hat{\gamma}/(c_1 r_r)\rceil$, which is finite and does not depend on $x_0$, we have $\hat{J}_{k-1}(x_{t_\mathrm{f}}) \leq r_2$. Combining the above statements, we can conclude that any trajectory of the closed-loop system with $\hat{\pi}^\mathrm{ex}_\mathrm{proj}(\cdot)$ starting from $X_\mathrm{in}$ will reach $\mathcal{B}_{r_2}$ in less than $N+t_\mathrm{f}$ time steps. Finally, the positive invariance of $\mathcal{B}_{r_2}$ is straightforward if $a^*_2 \leq 0$, since we have $\hat{J}_{k-1}(x) \leq r_2  \Rightarrow  \hat{J}_{k-1}(f_{\mathrm{PWA}}(x, \hat{\pi}^\mathrm{ex}_\mathrm{proj}(x))) \leq r_2$ from \eqref{verify2}. This completes the proof of Theorem \ref{theorem4}.

\bibliographystyle{IEEEtran}
\bibliography{IEEEexample}
%
%
%

%\newpage
% your own custom macro containing the $\backslash${\tt{includegraphics}} command to make things
%\bf{If you include a photo:}\vspace{-33pt}
%\begin{IEEEbiography}[{\includegraphics[width=1in,height=1.25in,clip,keepaspectratio]{fig1}}]{Michael Shell}
%Use $\backslash${\tt{begin\{IEEEbiography\}}} and then for the 1st argument use $\backslash${\tt{includegraphics}} to declare and link the author photo.
%Use the author name as the 3rd argument followed by the biography text.
%\end{IEEEbiography}

\vfill

\end{document}